\documentclass[preprint,aps,prd,showpacs,floatfix,nofootinbib]{revtex4-1}
\usepackage{times}
\usepackage{euscript}
\usepackage{amsthm}
\usepackage{graphicx}
\usepackage{amsmath}
\usepackage{amssymb}
\usepackage{amsfonts}
\usepackage[english]{babel}
\usepackage{epstopdf}
\usepackage{multirow,makecell,array}
\usepackage{mathtools}
\usepackage{epstopdf}
\usepackage{color}
\RequirePackage{bbm}

\newcommand*{\I}{ {\rm i} }

\newcommand{\mbeq}{\overset{!}{=}}

\definecolor{IVAN}{rgb}{0.0, 0.0, 0.7}
\definecolor{CHRISTIAN}{rgb}{0.8, 0.0, 0.0}
\definecolor{INFO}{rgb}{0.1, 0.8, 0.0}
\usepackage{ulem}

\definecolor{amendments}{rgb}{0.0, 0.0, 1.0}

\begin{document}
\thispagestyle{empty}

\title{Pair production in temporally and spatially oscillating fields}

\author{Ivan~A.~Aleksandrov}\email{i.aleksandrov@spbu.ru}
\affiliation{Department of Physics, St. Petersburg State University, 7/9 Universitetskaya Naberezhnaya, 199034 Saint Petersburg, Russia}
\affiliation{Ioffe Institute, Politekhnicheskaya str. 26, 194021 Saint Petersburg, Russia}
\author{Christian Kohlf\"urst}\email{c.kohlfuerst@hzdr.de}
\affiliation{Helmholtz-Zentrum Dresden-Rossendorf, Bautzner Landstra{\ss}e 400, 01328 Dresden, Germany
\vspace{0.5cm}}

\begin{abstract}

Electron-positron pair production for inhomogeneous electric and magnetic fields oscillating in space and time is investigated. By employing accurate numerical methods (Furry-picture quantization and quantum kinetic theory), final particle momentum spectra are calculated and analyzed in terms of effective models. Furthermore, criteria for the applicability of approximate methods are derived and discussed. In this context, special focus is placed on the local density approximation, where fields are assumed to be locally homogeneous in space. Eventually, we apply our findings to the multiphoton regime. Special emphasis is on the importance of linear momentum conservation and the effect of its absence in momentum spectra within approximations based on local homogeneity of the fields.

\end{abstract}

\maketitle
\section{Introduction}\label{sec:intro}

Quantum field theory has taught us to see the quantum vacuum as a key factor towards understanding fundamental physical processes. The reason is
that the quantum vacuum is far from being empty but can rather be described as a nonlinear medium with fluctuating virtual particles acting as mediators of interactions.
In quantum electrodynamics (QED), for example, it is perpetual creation and annihilation of virtual electrons, positrons, and photons that gives rise to new, staggering
phenomena such as the Sauter-Schwinger effect \cite{sauter_1931,euler_heisenberg, schwinger_1951}, vacuum birefringence \cite{Toll:1952, Baier, Baier2}, and light-by-light scattering \cite{Euler:1935zz, Karplus:1950zza, Karplus:1950zz, Weisskopf}. 

Especially in the context of strong-field QED, understanding these effective interactions and finding ways to probe vacuum
nonlinearities have driven the evolution of the research field ever since;
for reviews see Refs. \cite{Dittrich:1985yb, Dittrich:2000zu, Marklund:2008gj, Dunne:2008kc, Heinzl:2008an, dipiazza_rmp_2012, Dunne:2012vv, reviewB}. 
However, testing these conjectures experimentally has proved to be challenging. Due to the small cross sections
of direct light-by-light scattering, it takes extremely high field strengths in order to obtain any signal of underlying nonlinearities. As a consequence,
experiments have been limited to atomic fields and highly charged ions (see Refs.~\cite{Lee:1998hu, Akhmadaliev:2001ik, dEnterria:2013zqi, Aaboud:2017bwk}, where high-energy processes were discussed). New theoretical developments concerning spontaneous pair production in low-energy ion collisions have been recently reported in Ref.~\cite{maltsev_prl_2019} (see also references therein).

In the context of measuring nonlinear quantum electrodynamics the SLAC E144 experiment has to be mentioned. The setup consisting of a Terawatt laser and an electron beam of 46.6 GeV, which was guided through the laser beam, was the first experiment to observe nonlinear Compton scattering and, moreover, nonlinear Breit-Wheeler pair production \cite{Burke:1997ewA, Burke:1997ewB}. In the years following this seminal observation, newly built laser facilities have been recognized as an opportunity to advance the field even further, and thus a variety of new theoretical predictions regarding the feasibility of observing similarly astonishing phenomena have been published (see, e.g., Refs.~\cite{Ringwald:2001ib, PhysRevLett.101.130404, Heinzl:2006xc}).
In particular creation of matter through high-intensity
electric fields, the famous Schwinger effect, and the prospect of finding direct proof for quantum vacuum nonlinearities has gained renewed interest as
current laser technology is on the brink of investigating these effects in laboratory.

This development also has a huge impact on the theoretical aspect of the research field as experimental conditions are often far away from perfect theoretical settings, cf. Refs.~\cite{Hebenstreit:2011wk, Gies:2005bz, SemiClassA, Jiang:2014bwa, Hebenstreit:2011pm, abdukerim_plb_2013, linder_prd_2015, abdukerim_2017, aleksandrov_prd_2017_1, otto_epja_2018,Lv:2018wpn, aleksandrov_prd_2018,aleksandrov_prd_2017_2,blinne_prd_2014,schneider_jhep_2016,torgrimsson_prd_2018,wang_pra_2019,olugh_prd_2019,li_prd_2017,adorno_ijmpa_2017,granz_plb_2019} considering various temporal and spatial pulse shapes.
The possibility of working with an imperfect vacuum as well as having imprecise laser and detector
equipment plays a huge role in evaluating the chances for detecting, e.g., signal photons \cite{Lundstrom:2005za, Gies:2017ygp}.  
Hence, this article is mainly devoted to analyzing currently used numerical methods and approximations in studies on electron-positron pair production. To be more specific, we evaluate two computational techniques based on the Furry-picture quantization formalism and quantum kinetic approaches, respectively, in order to determine the strengths and weaknesses of the corresponding numerical procedures. Moreover, we test the applicability of the local dipole approximation, where spatial dependencies are treated as locally homogeneous. This is an important
step forward in finding a suitable method to actually perform calculations with respect to pair production for given laboratory conditions. 

For this work, we consider two counterpropagating laser beams which are focused at the same point creating a highly intense standing-wave pattern. Such a setup provides a 
perfect environment for testing various numerical methods because the temporal structure allows one to study Schwinger as well as multiphoton effects and the inhomogeneous spatial structure is ideal to see the implications of magnetic fields. Furthermore,
the computational techniques applied offer a lot of flexibility allowing us to further investigate the fundamental differences in purely time-dependent electric fields
and time-dependent, spatially inhomogeneous backgrounds.

The article is structured as follows. In Sec.~\ref{sec:field} we introduce the external laser field and discuss its characteristic traits. Sections~\ref{sec:furry},
\ref{sec:dhw}, and \ref{sec:LDA} are the basis of this paper as we individually introduce the different methods step by step. To be more specific, Sec.~\ref{sec:furry} is 
about the Furry-picture quantization, in Sec.~\ref{sec:dhw} a phase-space formalism is introduced, and in Sec.~\ref{sec:LDA} we discuss the local dipole approximation (LDA). In the largest section, Sec.~\ref{sec:results}, we
present and discuss our results on the basis of a comparison between the methods for short-pulsed fields, Sec. \ref{sec:results_short}, as well as for
many-cycle pulses, Sec.~\ref{sec:results_long}.
The last section, Sec.~\ref{sec:discussion}, contains the conclusions of our study.

Throughout the article, we use $\hbar=c=1$ and display observables in terms of the electron mass $m$. The electron charge is $e<0$.

\section{External field configuration}\label{sec:field}

One of our goals is to study the performance of different numerical methods regarding calculating the particle yield for spatially inhomogeneous fields within various setups.
Moreover, we are determined to inspect various multiphoton signatures in standing-wave patterns which approximate a scenario involving two counterpropagating laser pulses. Hence, we introduce a vector potential of the form
\begin{equation}
{\boldsymbol A} (t, z) = \begin{pmatrix} A_x (t, z) \\ 0 \\ 0 \end{pmatrix} = \frac{\varepsilon E_{\rm cr}}{\omega}\, \mathrm{exp}\bigg (-\frac{t^6}{\tau^6} \bigg ) \sin \omega t \cos k_z z \, {\boldsymbol e}_x. \label{eq:field_config}
\end{equation}
This expression represents a standing wave being linearly polarized along the $x$ direction, with a cosine profile along the $z$ axis and a super-Gaussian temporal envelope. Four individual field parameters allow for great flexibility: the peak field strength $\varepsilon$, the pulse length $\tau$, and the field frequencies $\omega$ and $k_z$. 
The critical field strength $E_{\rm cr}=m^2/|e|$ has been factored out for the sake of convenience. The electric and magnetic fields are derived from Eq.~(\ref{eq:field_config}),
\begin{eqnarray}
{\boldsymbol E} (t, z) &=& \begin{pmatrix} E_x (t, z) \\ 0 \\ 0 \end{pmatrix} = \begin{pmatrix} -\partial_t A_x (t, z) \\ 0 \\ 0 \end{pmatrix} = \notag \\ 
  &-&\varepsilon E_{\rm cr} \, \mathrm{exp}\bigg (\! -\frac{t^6}{\tau^6} \bigg ) \bigg [ \cos \omega t - \frac{6}{\omega \tau}\, \bigg (\frac{t}{\tau} \bigg )^5 \sin \omega t \bigg ] \cos k_z z \, {\boldsymbol e}_x, \label{eq:el_field} \\
{\boldsymbol B} (t, z) &=& \begin{pmatrix} 0 \\ B_y (t, z) \\ 0 \end{pmatrix} = \begin{pmatrix} 0 \\ \partial_z A_x (t, z) \\ 0 \end{pmatrix} =  -\varepsilon E_{\rm cr} \, \frac{k_z}{\omega}\, \mathrm{exp}\bigg (\! -\frac{t^6}{\tau^6} \bigg ) \sin \omega t \sin k_z z \, {\boldsymbol e}_y.\label{eq:mag_field}
\end{eqnarray}
In this way a single vector potential is sufficient in order to interpolate from the Schwinger to the multiphoton regime and from few-cycle to many-cycle pulses. In addition, all fields vanish at asymptotic times
as $A(t \to \pm \infty) \to 0$. As we will discuss in Sec. \ref{sec:results}, neither does $\omega=k_z$ hold in general nor do the parameters $\omega$ and $k_z$ always represent physical
quantities. Although Maxwell's equations lead to $|k_z|/\omega \mbeq 1$, our toy model allows for a gradual change of this ratio in order to switch on/off the magnetic field component 
and make the spatial variations more/less pronounced. In what follows, it will provide additional insights into the particle dynamics and its relation to the features of the pair production process.
Furthermore, for few-cycle pulses $\omega \tau \approx \mathcal O (1)$, a Fourier analysis of the pulse profiles does not yield one dominant field frequency. In many-cycle pulses $\omega \tau \gg 1$, the situation changes drastically in particular close to the critical frequency $\omega \sim m$. Then, it can be shown that $\omega$ indeed corresponds to the photon energy and $k_z$ represents the photon momentum.

\section{Furry-picture quantization}\label{sec:furry}

The general formalism of quantization within the Furry picture is described in detail in Ref.~\cite{fradkin_gitman_shvartsman}. We will first briefly discuss how the rigorous QED approach allows one to extract the necessary pair-production probabilities from the one-particle solutions of the Dirac equation including the interaction with a classical external field $A^\mu$. After that we will show how this approach can be implemented in the case of the standing-wave background~(\ref{eq:field_config}).

The external classical field is treated nonperturbatively, thus allowing one to study also nonperturbative pair production, e.g., the Schwinger mechanism in strong quasistatic fields. On the other hand, we will completely neglect the quantized part of the electromagnetic field, which gives rise to, e.g., loop corrections. Furthermore, neither photon emission~\cite{di_piazza_prd_2005, blaschke_prd_2011, otto_kaempfer_prd_2017, aleksandrov_photons, smolyansky_mpla_2020} nor backreaction effects~\cite{kluger, bloch} are taken into account. In short, our techniques are exact to zeroth order in the fine-structure constant.

The external field is assumed to vanish outside the time interval $t_\text{in} < t < t_\text{out}$, where in the case of the field configuration~(\ref{eq:field_config}), $t_\text{in/out} \to \mp \infty$. We introduce two complete sets of the one-particle Hamiltonian eigenfunctions at $t=t_\text{in}$ and $t=t_\text{out}$, respectively:
\begin{eqnarray}
\mathcal{H}_e (t_\text{in})\, {}_\pm \varphi_n (\boldsymbol{x}) &=& {}_\pm \varepsilon_n (t_\text{in})\, {}_\pm \varphi_n (\boldsymbol{x}),\label{eq:in_schroedinger}\\
\mathcal{H}_e (t_\text{out})\, {}^\pm \varphi_n (\boldsymbol{x}) &=& {}^\pm \varepsilon_n (t_\text{out}) \, {}^\pm \varphi_n (\boldsymbol{x}),
\label{eq:in_schroedinger_2}
\end{eqnarray}
where $\mathcal{H}_e (t) = \boldsymbol{\alpha} [-\I \boldsymbol{\nabla} - e \boldsymbol{A} (x)] + e A_0 (x) + \beta m$, and the eigenvalues denoted by plus (minus) are positive (negative). These sets are orthonormal and complete with respect to the usual inner product. In terms of these functions, the field operator can be decomposed as
\begin{eqnarray}
\psi (\boldsymbol{x}) &=& \sum_{n} \big [a_n(t_\text{in}) \, {}_+ \varphi_n (\boldsymbol{x}) + b^\dagger_n(t_\text{in}) \, {}_- \varphi_n (\boldsymbol{x})\big ],
\label{eq:field_decomposition_schroedinger}\\
\psi (\boldsymbol{x}) &=& \sum_{n} \big [a_n(t_\text{out}) \, {}^+ \varphi_n (\boldsymbol{x}) + b^\dagger_n(t_\text{out}) \, {}^- \varphi_n (\boldsymbol{x})\big ],
\label{eq:field_decomposition_schroedinger_2}
\end{eqnarray}
where we have introduced the electron (positron) creation and annihilation operators $a^\dagger_n$ ($b^\dagger_n$) and $a_n$ ($b_n$), respectively. These operators satisfy the usual anticommutation relations. The Dirac-field Hamiltonian $H_e (t)= \int \! \psi^\dagger (\boldsymbol{x}) \mathcal{H}_e (t) \psi (\boldsymbol{x}) {\rm d} \boldsymbol{x}$ is then diagonalized at time instants $t_\text{in}$ and $t_\text{out}$.

We will turn to the Heisenberg representation which is achieved by the unitary evolution operator
\begin{equation}
U_e (t, t') = T~\mathrm{exp} \Bigg ( -\I \int \limits_{t'}^t H_e (\tau) {\rm d} \tau \Bigg ).
\label{eq:evolution_operator_e}
\end{equation}
We perform a transformation by means of the operator $U_e (0, t)$, so the field operator gains a temporal dependence: 
\begin{equation}
\psi(t,\boldsymbol{x}) \equiv \psi(x) = U_e (0, t) \psi(\boldsymbol{x}) U^\dagger_e (0, t).
\label{eq:psi_heis}
\end{equation}
The creation and annihilation operators are transformed according to
\begin{eqnarray}
a_n(\text{in}) &=& U_e (0, t_\text{in}) a_n (t_\text{in}) U^\dagger_e (0, t_\text{in}),
\label{eq:in_out_operators_1}\\
a_n(\text{out}) &=& U_e (0, t_\text{out}) a_n(t_\text{out}) U^\dagger_e (0, t_\text{out}).
\label{eq:in_out_operators_2}
\end{eqnarray}
The other creation/annihilation operators with indices (in) and (out) are defined similarly. The anticommutation relations match those taking place in the Schr\"odinger picture. The {\it in} ({\it out}) vacuum in the Schr\"odinger and Heisenberg representation is denoted by $|0,t_\text{in} \rangle$ ($|0,t_\text{out} \rangle$) and $|0,\text{in} \rangle$ ($|0,\text{out} \rangle$), respectively, i.e., $|0,\text{in} \rangle = U_e (0, t_\text{in}) |0,t_\text{in} \rangle$ and $|0,\text{out} \rangle = U_e (0, t_\text{out}) |0,t_\text{out} \rangle$.

One can demonstrate that the evolution of the field operator $\psi (x)$ is governed by the equation $\I \partial_t \, \psi(x) = \mathcal{H}_e(t) \psi(x)$, i.e., it is a solution to the Dirac equation similar to that for the time-dependent one-particle solutions. From this, it follows that the field operator can be represented as
\begin{equation}
\psi (x) = \sum_{n} \big [a_n(\text{in}) \, {}_+ \varphi_n (x) + b^\dagger_n(\text{in}) \, {}_- \varphi_n (x)\big ], \label{eq:psi_x_in}
\end{equation}
where ${}_+ \varphi_n (x)$ and ${}_- \varphi_n (x)$ are the so-called {\it in} solutions of the Dirac equation evolved from the corresponding functions~(\ref{eq:in_schroedinger}). One can also show that
\begin{equation}
\psi (x) = \sum_{n} \big [a_n(\text{out}) \, {}^+ \varphi_n (x) + b^\dagger_n(\text{out}) \, {}^- \varphi_n (x)\big ], \label{eq:psi_x_out}
\end{equation}
where ${}^+ \varphi_n (x)$ and ${}^- \varphi_n (x)$ are the {\it out} solutions, which coincide at $t=t_\text{out}$ with the eigenfunctions~(\ref{eq:in_schroedinger_2}).

If the Schr\"odinger field operator $\psi (\boldsymbol{x})$ is known, the Heisenberg operator $\psi (t_\text{out}, \boldsymbol{x})$ can be constructed in two different ways. First, one can perform the transformation~(\ref{eq:psi_heis}) at $t=t_\text{out}$. Second, one can evolve the Heisenberg operator $\psi (t_\text{in}, \boldsymbol{x})$ by means of the one-particle propagator. This allows one to express the {\it out} operators in terms of the {\it in} operators according to
\begin{eqnarray}
a_n (\text{out}) &=& \sum_k \big [ a_k (\text{in}) G({}^+|{}_+)_{nk} + b^\dagger_k (\text{in}) G({}^+|{}_-)_{nk} \big ],\label{eq:out_in_1}\\
b_n (\text{out}) &=& \sum_k \big [ a^\dagger_k (\text{in}) G({}_+|{}^-)_{kn} + b_k (\text{in}) G({}_-|{}^-)_{kn} \big ],\label{eq:out_in_2}
\end{eqnarray}
where
\begin{equation}
G({}_\zeta|{}^\kappa)_{nk} = ({}_\zeta \varphi_n, {}^\kappa \varphi_k),\quad G({}^\zeta|{}_\kappa)_{nk} = ({}^\zeta \varphi_n, {}_\kappa \varphi_k), \quad \zeta, \kappa = \pm.
\label{eq:G_inner_product}
\end{equation}
Note that these inner products do not depend on time.

We can now evaluate the number density of electrons produced. The starting point is the following expression
\begin{equation}
n_m  (e^-) = \langle 0, t_\text{in} | U^\dagger_e (t_\text{out}, t_\text{in}) a^\dagger_m (t_\text{out}) a_m (t_\text{out}) U_e (t_\text{out}, t_\text{in}) | 0, t_\text{in} \rangle, \label{eq:number_density_start}
\end{equation}
which is obvious in the Schr\"odinger representation. Using the relation $U_e (t_\text{out}, t_\text{in}) = U_e (t_\text{out}, 0) U_e (0, t_\text{in})$ and inserting the identity operator $\mathbbm{1} = U_e (t_\text{out}, 0) U^\dagger_e (t_\text{out}, 0)$ between $a^\dagger_m (t_\text{out})$ and $a_m (t_\text{out})$, we find
\begin{equation}
n_m (e^-) = \langle 0,\text{in} | a^\dagger_m (\text{out}) a_m (\text{out}) |0,\text{in}\rangle.
\label{eq:number_density_heis}
\end{equation}
With the aid of Eqs.~(\ref{eq:out_in_1}) and (\ref{eq:out_in_2}), one obtains
\begin{equation}
n_m (e^-) = \sum_n G({}^+|{}_-)_{mn} G({}_-|{}^+)_{nm} = \{ G({}^+|{}_-) G({}_-|{}^+) \}_{mm}.
\label{eq:number_density_zeroth_order_G}
\end{equation}
The analogous expressions for the number density of positrons have the form
\begin{eqnarray}
n_m (e^+) &=& \langle 0,\text{in} | b^\dagger_m (\text{out}) b_m (\text{out}) |0,\text{in}\rangle,\\ \label{eq:number_density_heis_pos}
n_m (e^+) &=& \sum_n G({}^-|{}_+)_{mn} G({}_+|{}^-)_{nm} = \{ G({}^-|{}_+) G({}_+|{}^-) \}_{mm}. \label{eq:number_density_zeroth_order_G_pos}
\end{eqnarray}
It turns out that the full information about the number of particles and their spectrum (to zeroth order in the fine-structure constant) can be extracted from the ({\it in} and {\it out}) one-particle solutions of the Dirac equation incorporating the interaction with the external background. 

We will now discuss how one can employ this approach in the case of a standing wave~(\ref{eq:field_config}), which can be represented as $A_x (t,z) = Q(t) \cos k_z z$. Since the external field depends only on $t$ and $z$, the {\it in} and {\it out} solutions of the Dirac equation have the form
\begin{eqnarray}
{}^+ \varphi_{\boldsymbol{p},s} (x) = (2\pi)^{-3/2} \, \mathrm{e}^{\I \boldsymbol{p} \boldsymbol{x}} \, {}^+ \chi_{\boldsymbol{p},s} (t, z), &\quad \quad & {}_+ \varphi_{\boldsymbol{p},s} (x) = (2\pi)^{-3/2} \, \mathrm{e}^{\I \boldsymbol{p} \boldsymbol{x}} \, {}_+ \chi_{\boldsymbol{p},s} (t, z),\label{eq:phi_plus_chi_sw}\\
{}^- \varphi_{\boldsymbol{p},s} (x) = (2\pi)^{-3/2} \, \mathrm{e}^{-\I \boldsymbol{p} \boldsymbol{x}} \, {}^- \chi_{\boldsymbol{p},s} (t, z), &\quad \quad & {}_- \varphi_{\boldsymbol{p},s} (x) = (2\pi)^{-3/2} \, \mathrm{e}^{-\I \boldsymbol{p} \boldsymbol{x}} \, {}_- \chi_{\boldsymbol{p},s} (t, z).\label{eq:phi_minus_chi_sw}
\end{eqnarray}
The time-dependent functions $\chi$ satisfy the following conditions:
\begin{eqnarray}
{}^+ \chi_{\boldsymbol{p},s} (t\geq t_\text{out}, z) = \mathrm{e}^{-\I p^0 (t-t_\text{out})} \, u_{\boldsymbol{p},s}, &\quad \quad & {}_+ \chi_{\boldsymbol{p},s} (t\leq t_\text{in}, z) = \mathrm{e}^{-\I p^0 (t-t_\text{in})} \, u_{\boldsymbol{p},s},\label{eq:chi_initial_cond_1_sw}\\
{}^- \chi_{\boldsymbol{p},s} (t\geq t_\text{out}, z) = \mathrm{e}^{\I p^0 (t-t_\text{out})} \, v_{-\boldsymbol{p},s}, &\quad \quad & {}_- \chi_{\boldsymbol{p},s} (t\leq t_\text{in}, z) = \mathrm{e}^{\I p^0 (t-t_\text{in})} \, v_{-\boldsymbol{p},s},
\label{eq:chi_initial_cond_2_sw}
\end{eqnarray}
where $p^0 = \sqrt{m^2 + \boldsymbol{p}^2}$, and the bispinors $u_{\boldsymbol{p},s}$ and $v_{\boldsymbol{p},s}$ form an orthonormal and complete set (for given $\boldsymbol{p}$). Since the external field is monochromatic as a function of $z$, one can represent the functions ${}^\zeta \chi_{\boldsymbol{p},s} (t, z)$ as
\begin{equation}
{}^\zeta \chi_{\boldsymbol{p},s} (t, z) = \sum_{j=-\infty}^{+\infty} {}^\zeta w^j_{\boldsymbol{p},s} (t) \mathrm{e}^{\I \zeta k_z j z},
\label{eq:fourier_chi}
\end{equation}
where $\zeta = \pm$. Similar series are introduced for ${}_\zeta \chi_{\boldsymbol{p},s} (t, z)$. In terms of the time-dependent Fourier coefficients ${}^\zeta w^j_{\boldsymbol{p},s} (t)$, the Dirac equation takes the following form:
\begin{equation}
\I \, {}^\zeta \dot{w}^j_{\boldsymbol{p},s} (t) = \big [\zeta \, \boldsymbol{\alpha} \cdot ( \boldsymbol{p} + j \boldsymbol{K} ) +  \beta m \big ] \, {}^\zeta w^j_{\boldsymbol{p},s} (t) - \frac{e}{2} Q(t) \alpha_x \big [ {}^\zeta w^{j-1}_{\boldsymbol{p},s} (t) + {}^\zeta w^{j+1}_{\boldsymbol{p},s} (t) \big ],
\label{eq:w_eq_gen}
\end{equation}
where $\boldsymbol{K} = k_z \boldsymbol{e}_z$ and $j\in \mathbb{Z}$, so one has to solve an infinite system of ordinary differential equations (in practical calculations, it can, of course, be truncated at sufficiently large $|j|$). The same holds true for the {\it in} functions ${}_\zeta w^j_{\boldsymbol{p},s} (t)$.

The $G$ matrices can be evaluated according to
\begin{eqnarray}
G({}_+|{}^+)_{\boldsymbol{p},s; \, \boldsymbol{p}',s'} &=& \sum_l \delta (\boldsymbol{p} - \boldsymbol{p}' - l \boldsymbol{K}) \, g({}_+|{}^+)_{\boldsymbol{p},s,l,s'},\label{eq:G_g_1_sw}\\
G({}_+|{}^-)_{\boldsymbol{p},s; \, \boldsymbol{p}',s'} &=& \sum_l \delta (\boldsymbol{p} + \boldsymbol{p}' + l \boldsymbol{K}) \, g({}_+|{}^-)_{\boldsymbol{p},s,l,s'},\label{eq:G_g_2_sw}\\
G({}_-|{}^+)_{\boldsymbol{p},s; \, \boldsymbol{p}',s'} &=& \sum_l \delta (\boldsymbol{p} + \boldsymbol{p}' + l \boldsymbol{K}) \, g({}_-|{}^+)_{\boldsymbol{p},s,l,s'},\label{eq:G_g_3_sw}\\
G({}_-|{}^-)_{\boldsymbol{p},s; \, \boldsymbol{p}',s'} &=& \sum_l \delta (\boldsymbol{p} - \boldsymbol{p}' - l \boldsymbol{K}) \, g({}_-|{}^-)_{\boldsymbol{p},s,l,s'}, \label{eq:G_g_4_sw}
\end{eqnarray}
where
\begin{eqnarray}
g({}_+|{}^+)_{\boldsymbol{p},s,l,s'} &=& \sum_j [{}_+ w^j_{\boldsymbol{p}, s}(t)]^\dagger \, {}^+ w^{l+j}_{\boldsymbol{p} - l \boldsymbol{K},s'} (t),\label{eq:g_1_sw}\\
g({}_+|{}^-)_{\boldsymbol{p},s,l,s'} &=& \sum_j [{}_+ w^j_{\boldsymbol{p}, s}(t)]^\dagger \, {}^- w^{l-j}_{-\boldsymbol{p} - l \boldsymbol{K},s'} (t),\label{eq:g_2_sw}\\
g({}_-|{}^+)_{\boldsymbol{p},s,l,s'} &=& \sum_j [{}_- w^j_{\boldsymbol{p}, s}(t)]^\dagger \, {}^+ w^{l-j}_{-\boldsymbol{p} - l \boldsymbol{K},s'} (t),\label{eq:g_3_sw}\\
g({}_-|{}^-)_{\boldsymbol{p},s,l,s'} &=& \sum_j [{}_- w^j_{\boldsymbol{p}, s}(t)]^\dagger \, {}^- w^{l+j}_{\boldsymbol{p} - l \boldsymbol{K},s'} (t).\label{eq:g_4_sw}
\end{eqnarray}
The number density of electrons produced reads
\begin{equation}
n_{\boldsymbol{p},s} \equiv \frac{(2\pi )^3}{V} \frac{dN^{(-)}_{\boldsymbol{p},s}}{d\boldsymbol{p}} = \sum_l \sum_{s'} |g({}_-|{}^+)_{-\boldsymbol{p}-l\boldsymbol{K},s',l,s}|^2 = \sum_l \sum_{s'} |v^\dagger_{\boldsymbol{p}+l\boldsymbol{K},s'} {}^+ w^l_{\boldsymbol{p},s} (t_\text{in})|^2.
\label{eq:num_el_g_sw}
\end{equation}
Evolving the Fourier coefficients ${}^+ w^j_{\boldsymbol{p}, s} (t)$ backwards in time according to the system~(\ref{eq:w_eq_gen}), one can obtain the electron number density for given quantum numbers $\boldsymbol{p}$ and $s$. It turns out that in the case of a linearly polarized standing wave~(\ref{eq:field_config}), the number density does not depend on $s$, $n_{\boldsymbol{p},s} = n_{\boldsymbol{p}}$ (see, e.g., Ref.~\cite{woellert_2015}). The individual contributions for a given $l$ in Eq.~(\ref{eq:num_el_g_sw}) can be interpreted as separate channels of pair production by absorbing external-field quanta of total momentum $l \boldsymbol{K}$. This aspect will be addressed in Sec.~\ref{sec:results_long}.

Finally, we point out that our numerical technique was successfully employed in a number of previous studies~\cite{aleksandrov_prd_2016, aleksandrov_prd_2017_2, aleksandrov_prd_2018, aleksandrov_prd_2019, aleksandrov_photons}.

\section{Dirac-Heisenberg-Wigner formalism}\label{sec:dhw}

In heavy contrast to the Furry-picture formalism, Sec. \ref{sec:furry}, where one solves for the wave functions of one-particle
solutions to the Dirac equation, kinetic theories are formulated on the basis of transport properties of a quantum plasma. 
The advantage of such a description is mainly given by the fact that the governing set of equations of motion only have to be calculated once in order to determine the full particle spectrum. The downside of this approach is the presence of nonlocal operators, required
to properly account for dynamical pair production. Furthermore, the Dirac-Heisenberg-Wigner (DHW) formalism accesses the full particle phase space, a speciality of kinetic approaches, that can become numerically challenging.

Similarly to Sec. \ref{sec:furry} for the Furry-picture formalism, we will only state the key points in the derivation of the DHW approach,
cf. Ref. \cite{Vasak:1987umA} for a complete derivation of the governing equations of motion or Refs. \cite{Vasak:1987umB, BB} for a more in-depth look at the features of the formalism.

We begin the derivation of the governing quantum kinetic equations of motion by stating the QED Lagrangian
\begin{equation}
{\mathcal L} \left( \Psi, \bar{\Psi}, A \right) = \\
\frac{1}{2} \left( \I \bar{\Psi} \gamma^{\mu} \mathcal{D}_{\mu} \Psi - \I \bar{\Psi} \mathcal{D}_{\mu}^{\dag} \gamma^{\mu} \Psi \right) 
 -m \bar{\Psi} \Psi - \frac{1}{4} F_{\mu \nu} F^{\mu \nu}, \label{equ:Lag}
\end{equation}
and, consequently, the Dirac equation
\begin{eqnarray}
  \left(\I \gamma^{\mu} \partial_{\mu} - e \gamma^{\mu} A_{\mu} - m \right) \Psi &=& 0, \label{equ:Dir1} \\
  \bar{\Psi} \left(\I \overset{\leftharpoonup} {\partial_{\mu}} \gamma^{\mu} + e \gamma^{\mu} A_{\mu} + m \right) &=& 0, \label{equ:Dir2}
\end{eqnarray}
where $\mathcal{D}_{\mu} = \left( \partial_{\mu} +\I e A_{\mu} \right)$ and $\mathcal{D}_{\mu}^{\dag} = \left( \overset{\leftharpoonup} {\partial_{\mu}} -\I e A_{\mu} \right)$ 
are the covariant derivatives with a vector potential $A_\mu$ that vanishes at asymptotic times, and $\gamma^\mu$ 
are the gamma matrices. 
The fundamental quantity within the DHW approach is given by the gauge-invariant density operator
\begin{equation}
 \hat {\mathcal C}_{\alpha \beta} \left( r , s \right) = \mathcal U \left(A,r,s 
\right) \ \left[ \bar {\Psi}_\beta \left( r - s/2 \right), {\Psi}_\alpha \left( r + 
s/2 \right) \right],
\end{equation}
with the center-of-mass coordinate $r$, the relative coordinate $s$, and the Wilson line factor
\begin{equation}
 \mathcal U \left(A,r,s \right) = \exp \left( \mathrm{ie} \int_{-1/2}^{1/2} {\rm d} 
\xi \ A \left(r+ \xi s \right) \, s \right).
\end{equation}
In order to obtain a proper phase-space formalism, we perform a Fourier transform in $s$ leading to the covariant Wigner operator
\begin{equation}
 \hat{\mathcal W}_{\alpha \beta} \left( r , p \right) = \frac{1}{2} \int {\rm d}^4 s \ 
\mathrm{e}^{\mathrm{i} ps} \  \hat{\mathcal C}_{\alpha \beta} \left( r , s 
\right), \label{equ:W}
\end{equation}
which is properly defined in terms of four-position $r$ and four-momentum $p$ coordinates. Combining the Dirac and adjoint Dirac equations~\eqref{equ:Dir1}--\eqref{equ:Dir2} with 
derivatives of this operator yields two coupled operator equations
\begin{alignat}{3}
 & \left( \frac{1}{2} \hat D_{\mu}  - \I \hat P_{\mu}  \right) \gamma^{\mu} \hat{\mathcal W} \left( r , p \right) && = - &&\I m \hat{\mathcal W} \left( r , p \right), \label{equ:W1} \\
 & \left( \frac{1}{2} \hat D_{\mu}  + \I \hat P_{\mu}  \right) \hat{\mathcal W} \left( r , p \right) \gamma^{\mu}  && = &&\I m \hat{\mathcal W} \left( r , p \right), \label{equ:W2} 
\end{alignat}
where the derivatives $\partial_\mu$ are replaced by nonlocal, pseudodifferential operators 
\begin{alignat}{4}
 & \hat D_{\mu}  && = \partial_{\mu}^r - e &&\int_{-1/2}^{1/2} {\rm d} \xi \ && \hat F_{\mu \nu} \left( r - \I \xi \partial^p \right) \partial_p^{\nu}, \\
 & \hat P_{\mu}  && = p_{\mu} - \I e && \int_{-1/2}^{1/2} {\rm d} \xi \ \xi \ && \hat F_{\mu \nu} \left( r - \I \xi \partial^p \right) \partial_p^{\nu}.
\end{alignat}

In a crucial step towards obtaining a numerically feasible formalism based on distribution functions, we then take the vacuum
expectation value of Eqs.~\eqref{equ:W1}--\eqref{equ:W2}. At this point we introduce a Hartree-type approximation of the form
\begin{equation}
 \langle \Phi | \hat F^{\mu \nu} (r) | \Phi \rangle =  \langle \Phi | F^{\mu \nu} (r) + \mathcal{O} | \Phi \rangle \approx F^{\mu \nu} (r).
\end{equation}
This basically transforms the field-strength tensor $\hat F^{\mu \nu}$ to a fixed c-number valued function $F^{\mu \nu}$, i.e., we will perform calculations at tree level only neglecting the loop corrections and radiation processes as was also done in Sec.~\ref{sec:furry}. The key implication of this
approximation becomes apparent when considering terms of the form $\langle \hat F^{\mu \nu} \left( r \right) \ \hat{\mathcal C} \left( r , s \right) \rangle$,
which simplify tremendously
\begin{equation}
 \langle \Phi | \hat F^{\mu \nu} \left( r \right) \ \hat{\mathcal C} \left( r , s \right) | \Phi \rangle 
  \approx F^{\mu \nu} \left( r \right) \langle \Phi | \hat{\mathcal C} \left( r , s \right) | \Phi \rangle. \label{equ:Hart}
\end{equation}
The significance of Eq.~\eqref{equ:Hart} is given by the fact that we end up with a closed expression of only one-particle correlations 
instead of an infinite BBGKY hierarchy of $n$-particle correlation functions. 
This, in turn, allows us to obtain the equations of motion for the covariant Wigner function
\begin{equation}
 \mathcal W \left( r , p \right) = \langle \Phi | \hat{\mathcal W} \left( r , p \right) | \Phi \rangle.
\end{equation}
A decomposition into Dirac bilinears
\begin{equation}
\mathcal W  \left( r , p \right) = \frac{1}{4} \left( \mathbbm{1} \mathbbm{S} + \I \gamma_5 \mathbbm{P} + \gamma^{\mu} \mathbbm{V}_{\mu} + \gamma^{\mu} \gamma_5 \mathbbm{A}_{\mu} + \sigma^{\mu \nu} \mathbbm{T}_{\mu \nu} \right) \label{equ:wigner}
\end{equation}
turns the occurring matrix-valued equations into a set of equations for the
c-number valued Wigner coefficients $ \mathbbm S, \mathbbm P, \mathbbm V_\mu, \mathbbm A_\mu$, and $\mathbbm T_{\mu \nu}$. 
Up to this point everything is formulated covariantly, which, however, poses the problem
of having to know the solution to the governing equations of motion for all times. 
To avoid this, we project on equal times $\int \frac{{\rm d}p_0}{2 \pi}$ thus
essentially reformulating everything in terms of an initial-value problem. As all covariant Wigner coefficients 
are transformed analogously $\big[{\mathbbm w} \left( t, \boldsymbol{x} , \boldsymbol{p} \right) = \int \frac{{\rm d}p_0}{2 \pi} \mathbbm{W} \left( r , p \right)\big]$,
we eventually end up with a set of coupled integro-differential equations for 
the equal-time Wigner coefficients,
  \begin{alignat}{4}
    & D_t \mathbbm{s}     && && -2 \boldsymbol{\Pi} \cdot \mathbbm{t_1} &&= 0, \label{eq_3_1} \\
    & D_t \mathbbm{p} && && +2 \boldsymbol{\Pi} \cdot \mathbbm{t_2} &&= -2m\mathbbm{a}_0,  \\
    & D_t \mathbbm{v}_0 &&+ \boldsymbol{D} \cdot \mathbbm{v} && &&= 0,  \\
    & D_t \mathbbm{a}_0 &&+ \boldsymbol{D} \cdot \mathbbm{a} && &&= 2m\mathbbm{p},  \\    
    & D_t \mathbbm{v} &&+ \boldsymbol{D} \ \mathbbm{v}_0 && +2 \boldsymbol{\Pi} \times \mathbbm{a} &&= -2m\mathbbm{t_1},  \\    
    & D_t \mathbbm{a} &&+ \boldsymbol{D} \ \mathbbm{a}_0 && +2 \boldsymbol{\Pi} \times \mathbbm{v} &&= 0,  \\
    & D_t \mathbbm{t_1} &&+ \boldsymbol{D} \times \mathbbm{t_2} && +2 \boldsymbol{\Pi} \ \mathbbm{s} &&= 2m\mathbbm{v},  \\    
    & D_t \mathbbm{t_2} &&- \boldsymbol{D} \times \mathbbm{t_1} && -2 \boldsymbol{\Pi} \ \mathbbm{p} &&= 0.    \label{eqn1_4} 
  \end{alignat} 
The vectors $\mathbbm{t_1}$ and $\mathbbm{t_2}$ are defined as $\mathbbm{t_1} = 2 \mathbbm{t}^{i0} \boldsymbol{e}_i$ and $\mathbbm{t_2} = \epsilon_{ijk} \mathbbm{t}^{jk} \boldsymbol{e}_i$. The pseudodifferential operators are given by
  \begin{alignat}{6}
     & D_t && = \partial_t &&+ e &&\int {\rm d} \xi &&\boldsymbol{E} \left( \boldsymbol{x}+\I \xi \boldsymbol{\nabla}_p,t \right) && ~ \cdot \boldsymbol{\nabla}_p, \label{eqn2_1}  \\
     & \boldsymbol{D} && = \boldsymbol{\nabla}_x &&+ e &&\int {\rm d} \xi &&\boldsymbol{B} \left( \boldsymbol{x}+\I \xi \boldsymbol{\nabla}_p,t \right) &&\times \boldsymbol{\nabla}_p,  \label{eqn2_2} \\
     & \boldsymbol{\Pi} && = \boldsymbol{p} &&- \I e &&\int {\rm d} \xi \xi &&\boldsymbol{B} \left( \boldsymbol{x}+\I \xi \boldsymbol{\nabla}_p,t \right) &&\times \boldsymbol{\nabla}_p.  \label{eqn2_3}
  \end{alignat}    
To mimic the vacuum-to-matter formalism introduced in Sec.~\ref{sec:furry}, we further employ vacuum initial conditions
\begin{alignat}{3}
  \mathbbm{s}_{\rm vac} \left(\boldsymbol{p} \right) = -\frac{m}{\sqrt{m^2 +
\boldsymbol{p}^2}}, \quad && 
  \boldsymbol{\mathbbm{v}}_{\rm vac} \left(\boldsymbol{p} \right) = -\frac{
\boldsymbol{p}}{\sqrt{m^2 + \boldsymbol{p}^2}}. \label{equ:vac}
\end{alignat}
At this point we can already interpret the functions of the different Wigner coefficients \cite{BB}. Most notably, $ \mathbbm{s}$ gives the mass density,
$ \mathbbm{v}_0$ yields the charge density, and $ \boldsymbol{\mathbbm {v}}$ describes the current density.
Moreover, we can construct observables, valid when evaluated at asymptotic times, e.g., the particle distribution function
\begin{equation}
n \left( \boldsymbol{x}, \boldsymbol{p} \right) = \frac{m \left( \mathbbm{s}-\mathbbm{s}_{\rm vac} \right) + {\boldsymbol p} \cdot \left(
\boldsymbol{\mathbbm{v}}-\boldsymbol{\mathbbm{v}}_{\rm vac} \right)}{2 \sqrt{m^2+\boldsymbol{p}^2}}
 \label{equ:n}
\end{equation}
as well as the particle momentum spectrum
\begin{equation}
 n \left( \boldsymbol{p} \right) = \int {\rm d}^3 {\boldsymbol x} ~ n \left( \boldsymbol{x}, \boldsymbol{p} \right).
 \label{equ:nn}
\end{equation}
As we are employing periodic boundary conditions in $\boldsymbol{x}$ and are mainly interested in particle densities, the spatial integral in Eq.~\eqref{equ:nn} requires proper normalization, $\int_{-\pi/k_x}^{\pi/k_x} \frac{{\rm d}x}{2 \pi /k_x} \ \int_{-\pi/k_y}^{\pi/k_y} \frac{{\rm d}y}{2 \pi /k_y} \ \int_{-\pi/k_z}^{\pi/k_z} \frac{{\rm d}z}{2 \pi /k_z}$. 

\section{Local dipole approximation}\label{sec:LDA}

As exact computations (see note in \cite{comment_exact}), taking into account spatiotemporal inhomogeneities of complex external backgrounds may be extremely time consuming, it is important to employ approximate techniques or design effective models to advance computations. If, for example, the external field only slowly varies in space and time, one can assume it to be locally constant, which provides a huge advantage since the Heisenberg-Euler effective Lagrangian~\cite{euler_heisenberg} has a well-known nonperturbative closed-form expression in the case of a constant background. A general idea here is to calculate the necessary physical quantities in the presence of a constant external field and then perform integration over space and time by summing the individual (local) contributions. 

The locally constant field approximation (LCFA) is widely used in particle-in-cell (PIC) simulations, where ``pair production'' refers to a two-step process with a charged particle (electron) as initial condition (see, e.g., Refs.~\cite{elkina_2011, duclous_2011, gonoskov_pre_2015}). This electron is accelerated by the background field leading to the emission of photons via nonlinear Compton scattering. At the second stage, these photons decay into electron-positron pairs via the (non)linear Breit-Wheeler mechanism. To be more specific, the general pair production probability $W(\chi, f, g)$ can be stated in terms of three parameters: the quantum parameter $\chi \left(p, \boldsymbol{E}, \boldsymbol{B} \right) = \frac{e \sqrt{\left(F^{\mu \nu} p_\nu \right)^2 }}{m^3}$, with the particle momentum ${\bf p}$ and the electromagnetic field strength tensor $F^{\mu \nu}$, and the two Lorentz invariants $f=\frac{\boldsymbol{E} \cdot \boldsymbol{B}}{E_{\rm cr}}$ and $g=\frac{\boldsymbol{E}^2 - \boldsymbol{B}^2}{E_{\rm cr}^2}$. For subcritical fields $f,g \ll 1$ and relativistic particle energies, as is generally the case for PIC simulations, the majority of quantum effects is induced by particles crossing the background field. Hence, the probability can be replaced with $W(\chi, 0, 0)$ and the external field is treated as locally constant allowing one to use the explicit expressions derived in the case of constant crossed fields~\cite{nikishov_ritus, ritus_1985}. The corresponding expressions for the rate of the quantum processes in constant crossed fields can then be locally employed in PIC codes.

In the present study, we consider the process of vacuum particle production by a classical external field without any initial particles, i.e., the external field amplitude is sufficiently large to create particles in a one-stage process. In the context of this phenomenon the applicability of the LCFA is still not perfectly well understood especially when evaluating the momentum spectra of particles produced in the presence of general space-time-dependent fields. The main obstacle here is the fact that the particle created is being accelerated in the field, so that its dynamics can become overwhelmingly complicated in a general background. Furthermore, the locally constant field approximation disregards photon absorption effects entirely thus restricting its validity severely. To overcome these issues, we abandon the strict concept of locally homogeneous fields and discuss a local density approximation instead. 
Our suggestion is to treat the external field as locally constant in space incorporating the temporal dependence exactly. Since approximating the field by a spatially uniform time-dependent background is often referred to as the dipole approximation~\cite{comment_DA}, we will refer to this technique as the local dipole approximation (LDA). In what follows, we will discuss how the LDA is implemented within the methods described in Secs.~\ref{sec:furry} and \ref{sec:dhw}. 

\subsection{LDA in the Furry-picture approach}\label{subsec:furry_LDA}


The external field is approximated by a spatially homogeneous background whose temporal dependence coincides with that of the original field considered at a given position $\boldsymbol{x}$. In this case, we perform exact computations of the momentum spectra by solving the Dirac equation and employing the formalism described in Sec.~\ref{sec:furry}. By varying then $\boldsymbol{x}$ and summing the results over the spatial coordinates, one obtains approximate momentum distributions of particles created. This approach was discussed in detail in Ref.~\cite{aleksandrov_prd_2019}. Although it was already benchmarked against the exact predictions considering several different space-time-dependent field configurations~\cite{aleksandrov_prd_2019}, it was not assessed in the case of spatially oscillating fields. In the present study, we carry out the analysis of the LDA by applying it to a standing-wave background. In this case, instead of integrating over the whole $z$ axis, one should average the results over the spatial period, i.e., $z$ is being varied within the region $[0,~2\pi/k_z]$.

Within the LDA computations, the external field does not depend on the spatial coordinates, which means that the coordinate part of the one-particle solutions of the Dirac equation is always given by $\mathrm{exp} (\pm \I \boldsymbol{p}\boldsymbol{x})$, so one has to construct only the time-dependent bispinor part, which is a solution to Eq.~(\ref{eq:w_eq_gen}) for $j=0$ in the absence of the coupling terms involving $Q(t)$. These terms do not appear within the LDA as the momentum transfer does not take place here. Instead of an infinite ODE system, one now deals with only one four-component equation.

Let us point out several important features concerning the validity of the LDA. First, we note that such approximate calculations completely neglect the magnetic field component, so one may expect this approach to perform well only for sufficiently small $k_z$ and large $\omega$ since the characteristic ratio between the electric field strength~(\ref{eq:el_field}) and the magnetic field strength~(\ref{eq:mag_field}) is $k_z/\omega$. Moreover, the LDA approximates the electric field component as well, so one can expect that averaging the results over $z$ is reasonable only when the external field slowly varies in space, i.e., $k_z$ is again sufficiently small. Note that the LDA predictions, i.e., the average pair-production probabilities, do not depend on $k_z$ at all. 
Thus, there should be a characteristic value $k^{(\text{LDA})}_z$ allowing one to employ the LDA for $k_z \lesssim k^{(\text{LDA})}_z$. 
In Sec.~\ref{sec:results}, we will investigate how $k^{(\text{LDA})}_z$ depends on the external field parameters and reveal several important features of this dependence.

\subsection{LDA in the DHW formalism}\label{subsec:dhw_LDA}

Due to the fact that evaluating the pseudodifferential operators in Eqs.~\eqref{eqn2_1}--\eqref{eqn2_3} is numerically challenging, the LDA provides a good opportunity to significantly reduce numerical
costs. The phase-space formalism is inherently formulated as an initial value problem for transport properties, e.g., mass or charge density. Hence, it is impossible to perform independent calculations for separate mode functions. 

Within the LDA, however, the governing equations of motion simplify drastically, as a Taylor expansion in ${\boldsymbol x}$ of the pseudodifferential operators 
terminates at first order and, as the spatial coordinates $\boldsymbol{x}_i$ are fixed, the spatial derivative operators $\nabla_{\boldsymbol x}$, and with them the magnetic fields, vanish entirely leading to a much simpler system of equations,
cf. Ref.~\cite{Hebenstreit:2010vz}. 
To be more specific, within the LDA the set of partial differential equations decouples into a set of ordinary differential equations. The particle momentum spectrum is then obtained by simply summing the individual local density functions $n \left( \boldsymbol{x}_i, \boldsymbol{p} \right)$ over all instances of $\boldsymbol{x}_i$.
See Refs. \cite{Kluger:1992md, Smolyansky:1997fcA, kluger, Smolyansky:1997fcC} for more details on a phase-space formalism for purely time-dependent electric fields.

In all our simulations, the LDA results obtained within the Furry-picture approach and DHW formalism were found to be identical (see Appendix~\ref{sec:appendix} for more details).

\section{Numerical results}\label{sec:results}

We vary the external field parameters within the following domain: $0.1 \leq \varepsilon \leq 1.0$, $0.1m \leq \omega \leq m$, $0.1m \leq k_z \leq m$. In what follows, we will separately consider the cases of short and long laser pulses, i.e., we will employ small and large values of the pulse duration $\tau$.

\subsection{Short pulses}\label{sec:results_short}

In this subsection we choose $\tau = 5m^{-1}$ and vary $\varepsilon$, $k_z$, and $\omega$. For such a small value of the pulse duration, the number of carrier cycles $N \sim \omega \tau /(2\pi)$ is always less than $1$ in our computations, so the carrier frequency is not well localized in the Fourier spectrum of the pulse. 

\begin{figure}[t]
\center{\includegraphics[width=0.65\linewidth]{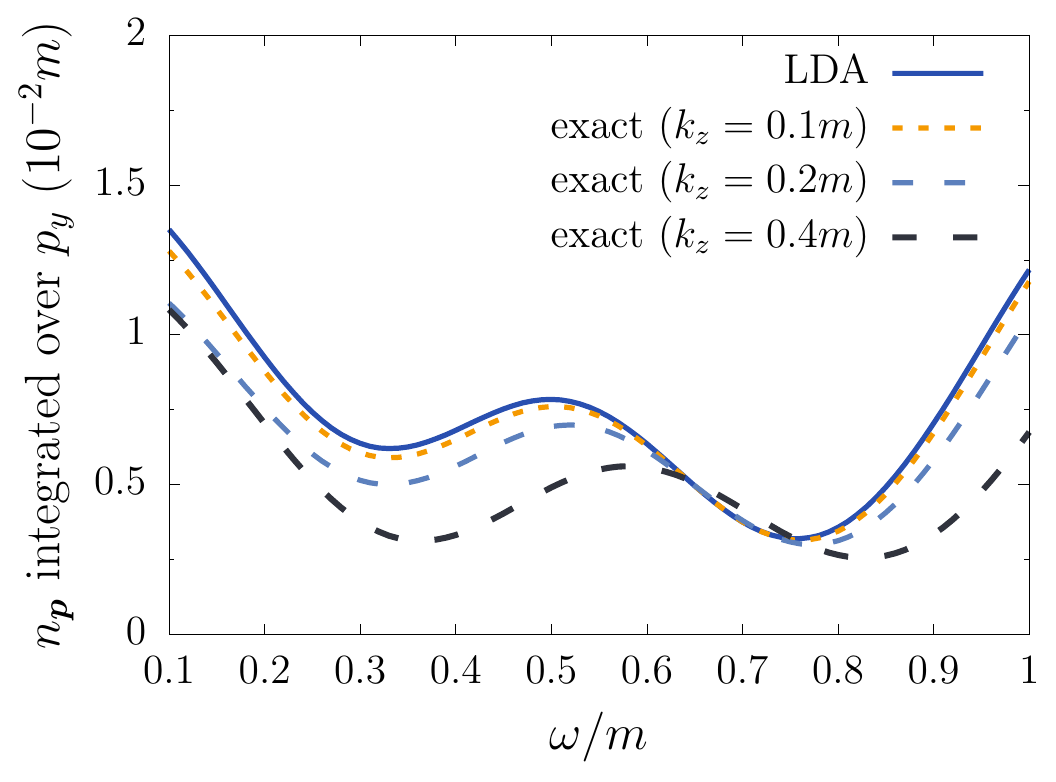}}
\caption{Electron number density $n_{\boldsymbol p}$ integrated over $p_y$ at $p_x = p_z = 0$ as a function of the carrier frequency $\omega$ for various values of $k_z$ ($\tau = 5 m^{-1}$, $\varepsilon = 0.2$).}
\label{fig:short_LDA_omega_02}
\end{figure}

\paragraph{Validity of the LDA.---} In order to determine $k^{(\text{LDA})}_z$, we compared the exact momentum distributions obtained by means of the techniques discussed in Secs.~\ref{sec:furry} and \ref{sec:dhw} to those predicted by the LDA by first varying $k_z$ for given $\varepsilon$ and $\omega$ and then turning to other values of these two parameters.
\begin{figure*}[t]
\center{\includegraphics[height=0.35\linewidth]{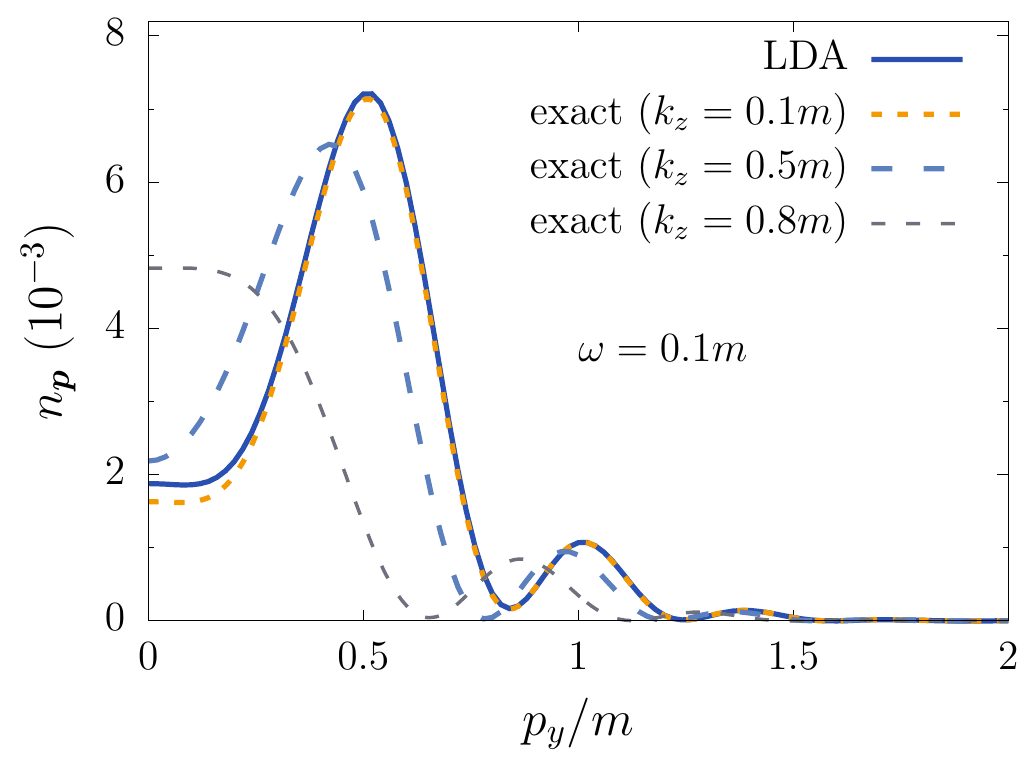}~~~~~\includegraphics[height=0.35\linewidth]{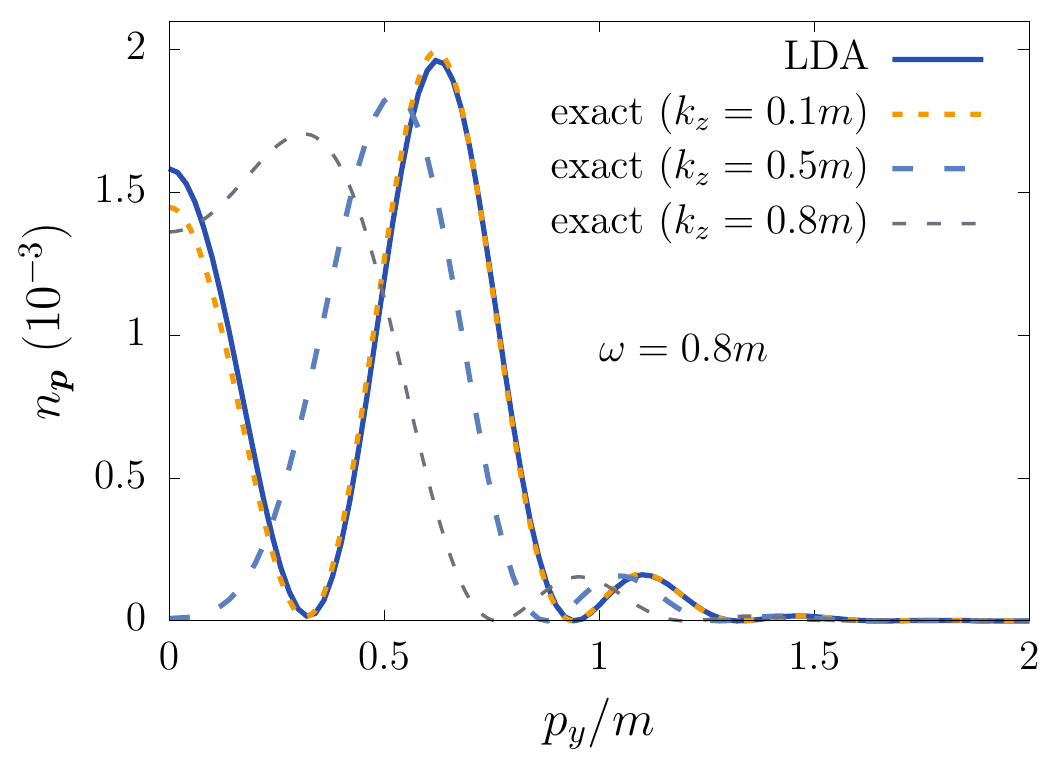}}
\caption{Momentum distributions computed by means of the LDA and calculated exactly for various values of $k_z$ ($p_x = p_z = 0$). The field parameters are $\tau = 5 m^{-1}$, $\varepsilon = 0.2$, $\omega = 0.1m$ (left) and $\omega = 0.8m$ (right).}
\label{fig:short_LDA_py_02}
\end{figure*}
In Fig.~\ref{fig:short_LDA_py_02} we present the momentum spectra along the magnetic field direction $y$ which were obtained exactly for various $k_z$ and with the aid of the LDA. We observe that the LDA indeed provides very accurate predictions in the case of small $k_z$ which become worse with increasing $k_z$. Note that the discrepancy between the exact and approximate results appears to be independent of $\omega$, i.e., it is fully determined by $k_z$. This fact is illustrated in Fig.~\ref{fig:short_LDA_py_02} and was confirmed in our computations employing other different field parameters. It means that $k^{(\text{LDA})}_z$ does not depend on $\omega$, which is no surprise as $\omega$ does not change much the form of the laser pulse for such short pulses. In order to make this point more evident, we evaluated the number density of electrons at $p_x = p_z = 0$ integrated over $p_y$ as a function of $\omega$ for various values of $k_z$ (see Fig.~\ref{fig:short_LDA_omega_02}). 
These curves provide a measure of the discrepancy between the LDA spectra and the exact results and indicate that the ratio $k_z/\omega$ does not play any important role for given $k_z$. The spectra computed within our study proved that this ratio is an irrelevant parameter in the case of such short laser pulses.

The fact that the validity of the LDA strongly depends on $k_z$ and does not depend on $\omega$ can also be accounted for by the requirement that the ``formation length'' $\ell = 2m/|e\varepsilon E_\text{cr}|$ must be much smaller than the spatial wavelength $\lambda = 2\pi/k_z$. In other words, the $e^+e^-$ pair must be formed within a sufficiently narrow spatial region. This condition is equivalent to $k_z/(\pi m\varepsilon) \ll 1$, which is indeed independent of $\omega$ and also explains why smaller values of $k_z$ are preferable. However, as will be seen below, this requirement is not the only criterion regarding the LDA justification.

Next we will examine how $k^{(\text{LDA})}_z$ depends on $\varepsilon$.
\begin{figure}[t]
\center{\includegraphics[width=0.65\linewidth]{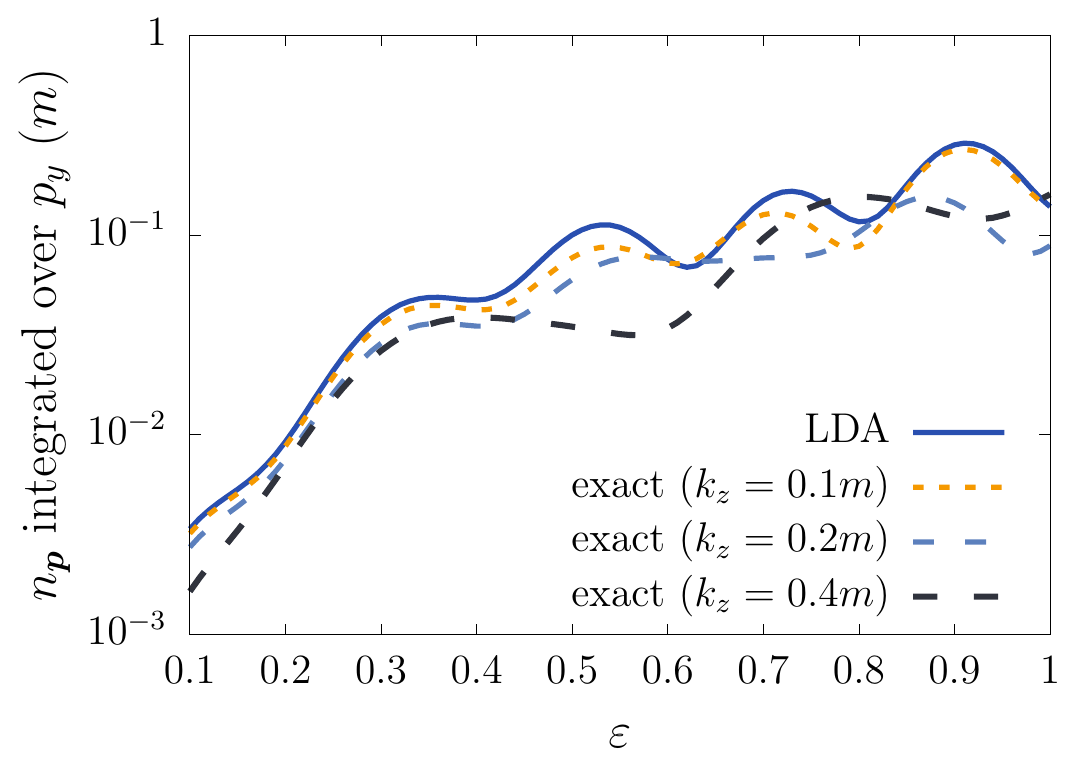}}
\caption{Electron number density $n_{\boldsymbol p}$ integrated over $p_y$ at $p_x = p_z = 0$ as a function of the field amplitude $\varepsilon$ for various $k_z$ ($\tau = 5 m^{-1}$, $\omega = 0.2 m$).}
\label{fig:short_LDA_eps_w_02}
\end{figure}
In Fig.~\ref{fig:short_LDA_eps_w_02} we present the number density of electrons integrated over $p_y$ for $p_x = p_z = 0$ as a function of $\varepsilon$ for various values of $k_z$ ($\omega = 0.2 m$). We observe that the discrepancy between the exact curves and the LDA predictions is not reduced for larger values of the field amplitude, so the results suggest that the concept of the formation length does not always provide a relevant parameter for justifying this kind of an approximate technique.

As was demonstrated in Ref.~\cite{aleksandrov_prd_2019}, in order to properly justify the LDA, one should also make sure that the classical particle trajectories are well localized within the spatial region where the external field does not change much. This condition arises due to the fact that the LDA treats different positions in space independently, and thus does not appropriately take into account the particle post-creation dynamics leading to inaccurate momentum spectra. This drawback becomes crucial when the particle trajectory stretches over a large part of the spatial period of the external standing-wave background. To carefully examine this issue, we solve the equations of motion in the case of a relativistic classical particle interacting with the standing electromagnetic wave of the form~(\ref{eq:field_config}). We assume that the particle is initially at rest [$\boldsymbol{p} (t_\text{in}) = 0$] at the position $z(t_\text{in}) = z_0$. Since the external field does not depend on $x$ and $y$, two momentum components have obvious temporal dependences:
\begin{eqnarray}
p_x (t) &=& e[A_x(t_\text{in}, z_0) - A_x(t, z(t))],\label{eq:class_px}\\
p_y (t) &=& 0.\label{eq:class_py}
\end{eqnarray}
In order to evaluate the function $z(t)$, we solve the following ODE system:
\begin{eqnarray}
\frac{dp_z}{dt} &=& \frac{e p_x(t) B_y(t, z(t))}{\sqrt{m^2 + p_x^2(t) + p_z^2(t)}},\label{eq:class_pz}\\
\frac{dz}{dt} &=& \frac{p_z (t)}{\sqrt{m^2 + p_x^2(t) + p_z^2(t)}}.\label{eq:class_z}
\end{eqnarray}
As the particle is likely to be produced in the vicinity of the electric field maxima $z = \pi n/k_z$ ($n \in \mathbb{Z}$), we choose $z_0$ to be close to the origin $z=0$. 
To take the magnetic field component into account properly, we use a small nonzero value $z_0 = \lambda/20 = \pi/(10k_z)$. We then propagate the solution $z(t)$ from $t=t_\text{in}$ to $t=1.5 \tau$, where the external field is about to vanish, and evaluate the ratio $\eta = |z(1.5\tau) - z_0|/\lambda$. In this context, $\eta$ is a measure for the particles' maximal displacement within the strong-field region of the background field. The criterion for the applicability of the LDA is given by $\eta \ll 1$.

\begin{figure*}[t]
\center{\includegraphics[height=0.36\linewidth]{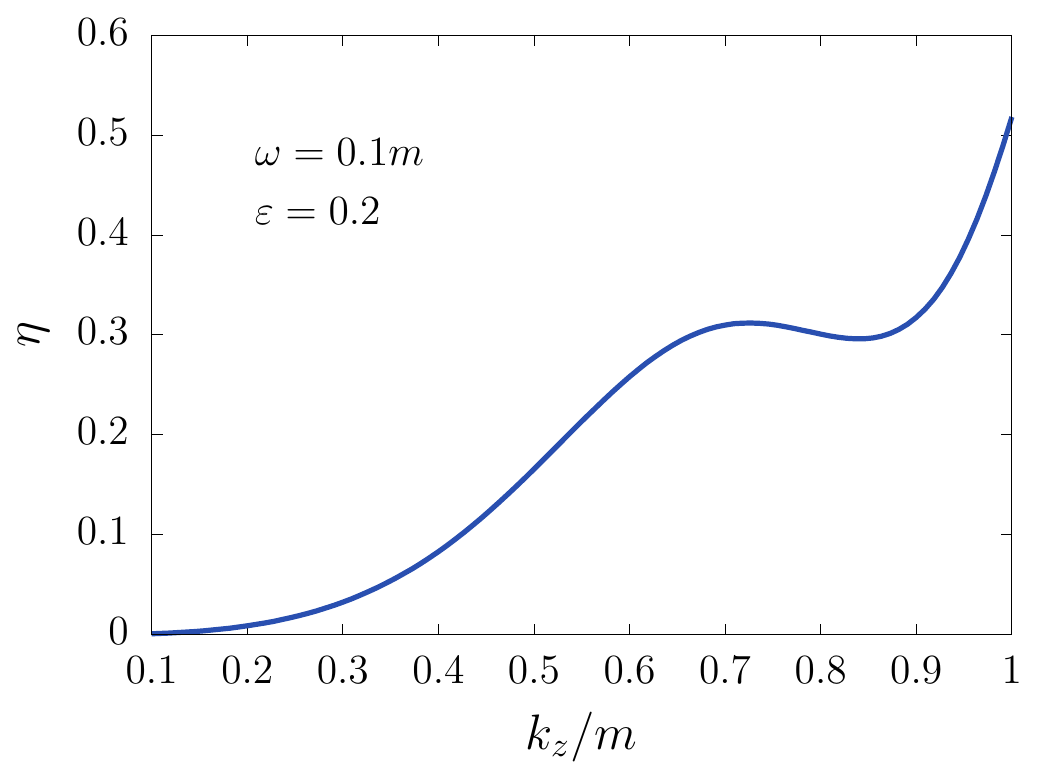}~~~~~\includegraphics[height=0.36\linewidth]{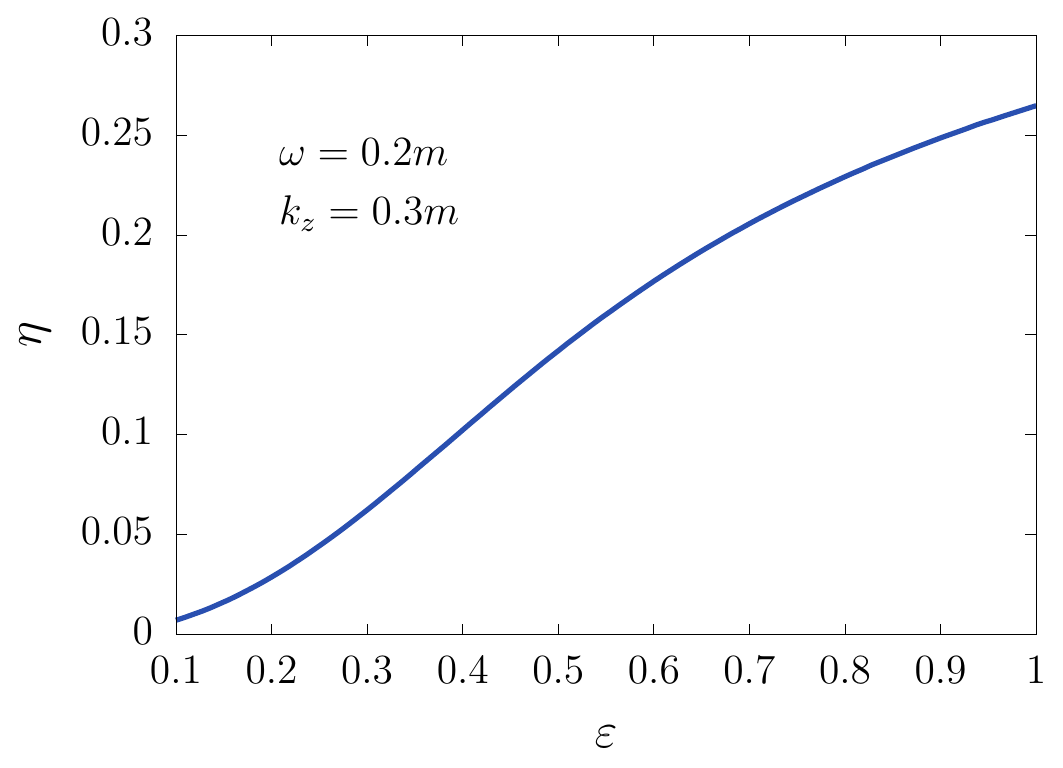}}
\caption{Parameter $\eta = |z(1.5\tau) - z_0|/\lambda$ as a function of $k_z$ for $\omega = 0.1m$ and $\varepsilon = 0.2$ (left) and its dependence on $\varepsilon$ for $\omega = 0.2m$ and $k_z = 0.3m$ (right). The pulse duration is $\tau = 5 m^{-1}$.}
\label{fig:short_LDA_eta}
\end{figure*}

\paragraph{Trajectory analysis.---} In Fig. \ref{fig:short_LDA_eta}
we find that only when $k_z$ is sufficiently small, particle positions fluctuate locally, i.e., $\eta \ll 1$. Starting with $k_z \approx 0.4 m$, the value of $\eta$ exceeds $0.1$, so we expect a noticeable discrepancy between the LDA and exact simulations. It is nicely illustrated in Fig. \ref{fig:short_LDA_py_02}, where calculations for $k_z=0.8m$ considerably deviate from the curves obtained through the LDA. More important, calculating $\eta$ as a function of $\varepsilon$, the discrepancies displayed in Fig. \ref{fig:short_LDA_eps_w_02} can be well understood by considering that for strong laser pulses ($\varepsilon \ge 0.4$) particle trajectories become poorly localized. On the other hand, for weak fields ($\varepsilon < 0.4$) the first criterion based on the particle formation length is not fulfilled as $\ell /\lambda > 0.24$. Accordingly, the LDA is inaccurate for fields where $k_z \ge 0.3m$ independent of $\varepsilon$. For slowly varying fields with $k_z \sim 0.1m$, on the other hand, $\eta$ remains small for all $\varepsilon \le 1$. Hence, applying the LDA and averaging over $z$ is completely justified according to this criterion. 

\paragraph{Formation length.---}
For the other criterion, the formation length $\ell$, one actually distinguishes in this example between strong and weak fields. For $\varepsilon \ge 0.2$ particles are created in a sufficiently confined space interval. For the latter, however, the ratio $\ell/\lambda$ increases with decreasing field strength and thus pair creation happens nonlocally. As for such weak fields the trajectories of the created particles are approximately stationary, averaging over $z$ still yields good agreement with the full calculations, cf. Fig.~\ref{fig:short_LDA_eps_w_02}. This example demonstrates that simple estimates involving the formation length do not necessarily yield a well-defined threshold $k^{(\text{LDA})}_z$.

As a similar argument can actually be made when $\eta$ is considered on its own, it shows that it is the combination of both criteria, $\ell/\lambda \ll 1$ and $\eta \ll 1$, that determines the validity of the LDA.

\paragraph{Magnetic fields.---}
Finally, let us note that the LDA leads to identical spectra in the $y$ and $z$ directions, whereas the exact results are anisotropic due to the presence of the magnetic field.
\begin{figure*}[t]
\center{\includegraphics[height=0.35\linewidth]{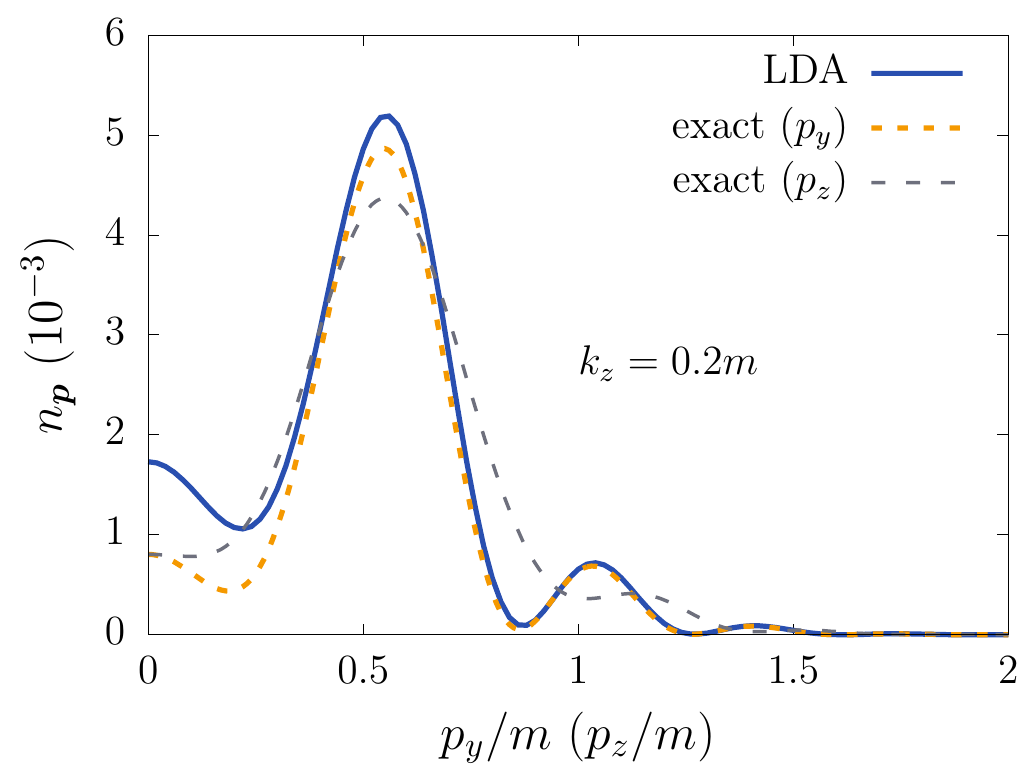}~~~~~\includegraphics[height=0.35\linewidth]{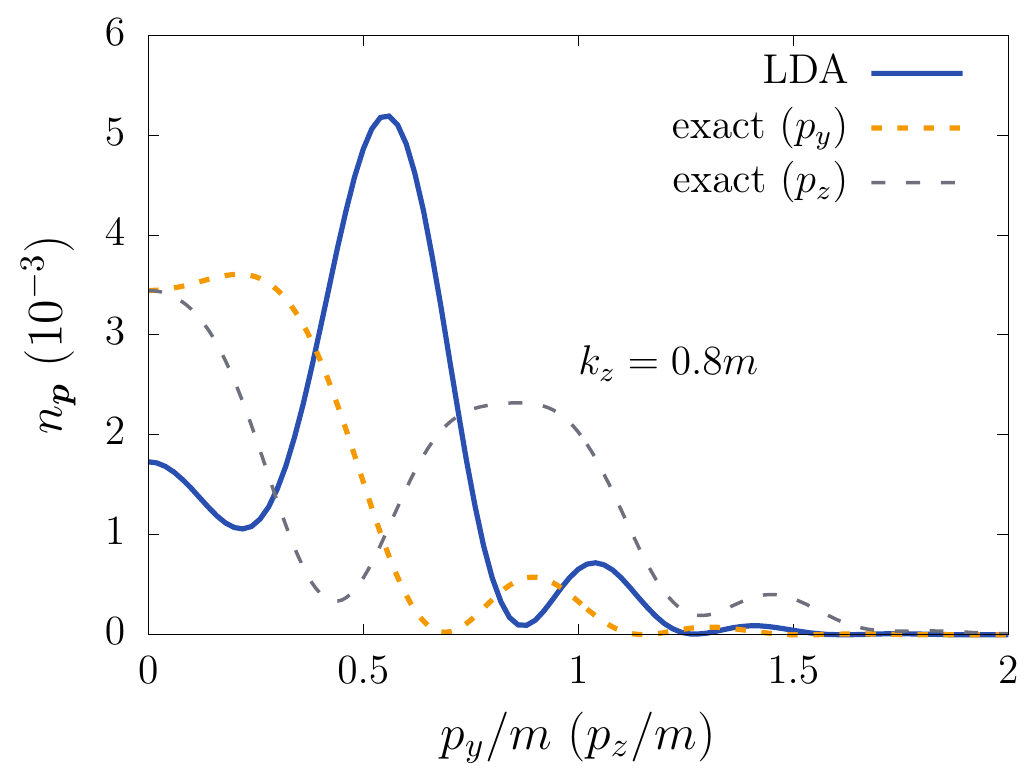}}
\caption{Momentum distributions with respect to $p_y$ and $p_z$ computed by means of the LDA and calculated exactly. The field parameters are $\tau = 5 m^{-1}$, $\varepsilon = 0.2$, $\omega = 0.2m$, and $k_z = 0.2 m$ (left) and $k_z = 0.8m$ (right).}
\label{fig:short_LDA_py_pz}
\end{figure*}
In Fig.~\ref{fig:short_LDA_py_pz} we display this anisotropy for two different values of $k_z$. We observe again that the accuracy of the LDA results strongly depends on $k_z$.

\subsection{Long-pulsed, high-frequency fields}
\label{sec:results_long}
      
External fields that exhibit a high frequency-duration product, \mbox{$\omega \tau \gg 1$}, open up an opportunity for multiphoton pair production. Especially for rapidly oscillating fields, photon absorption easily becomes the main source for particle creation. If we further assume that high-intensity light sources in the hard x-ray regime operate at a specific carrier frequency, setups involving collisions of multiple pulses give rise to distinctive and well-pronounced multiphoton signatures in the momentum spectrum. 

As the focus in this section of the manuscript is primarily on evaluating the importance of the incident photons' momenta, we rely again on a comparison between full-scale simulations and the LDA. In this way, we can perfectly isolate the impact that momentum transfer has on the particle spectra as in the LDA the ``energy packets'' do not carry any linear momentum.

To be more specific, we take advantage of the fact that we have access to various performant codes that excel in different areas. To gain an overview over the full particle spectrum, we solve the equations of motion within the DHW formalism and display the results in terms of 2D density plots. High-precision computations, which are required to perform an in-depth analysis, are done within the Furry-picture approach. In order to understand the outcome of these simulations in simple terms, we introduce a model based on momentum-energy conservation laws. This absorption model is capable of predicting the appearance of resonances in the particle distribution and thus perfectly suited to examine spectral signatures in the multiphoton regime. 

\begin{figure*}[t]
\center{\includegraphics[height=0.35\linewidth]{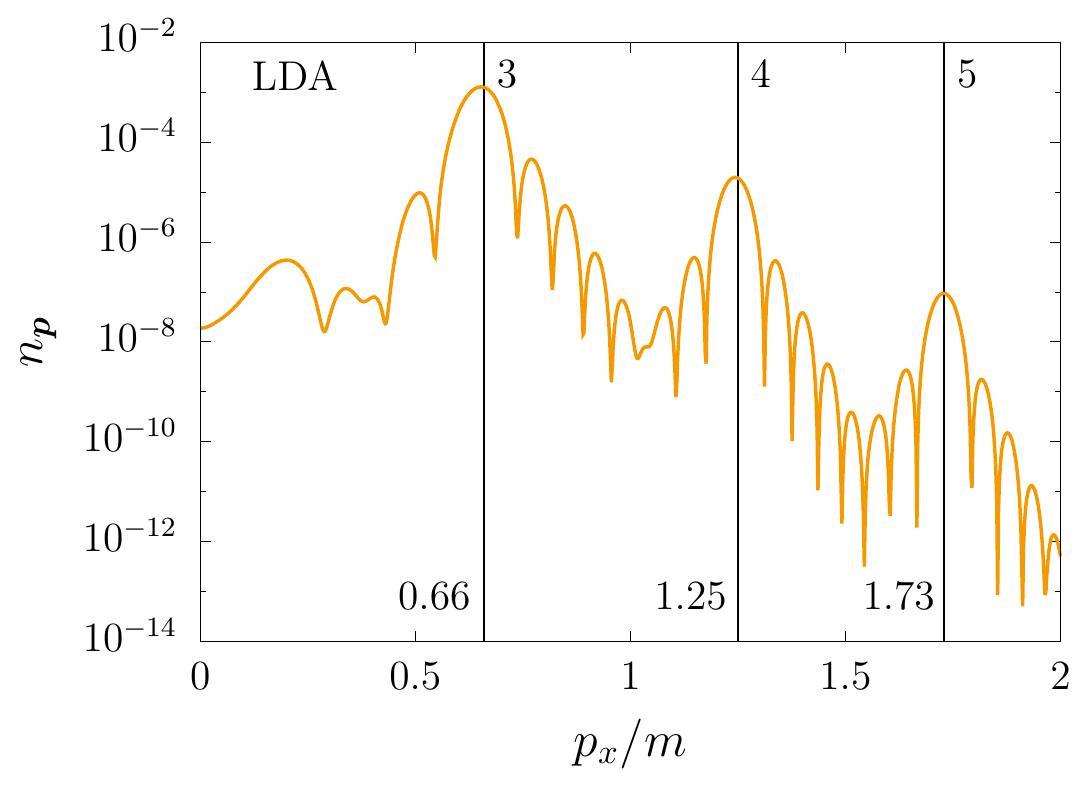}~~~~~\includegraphics[height=0.35\linewidth]{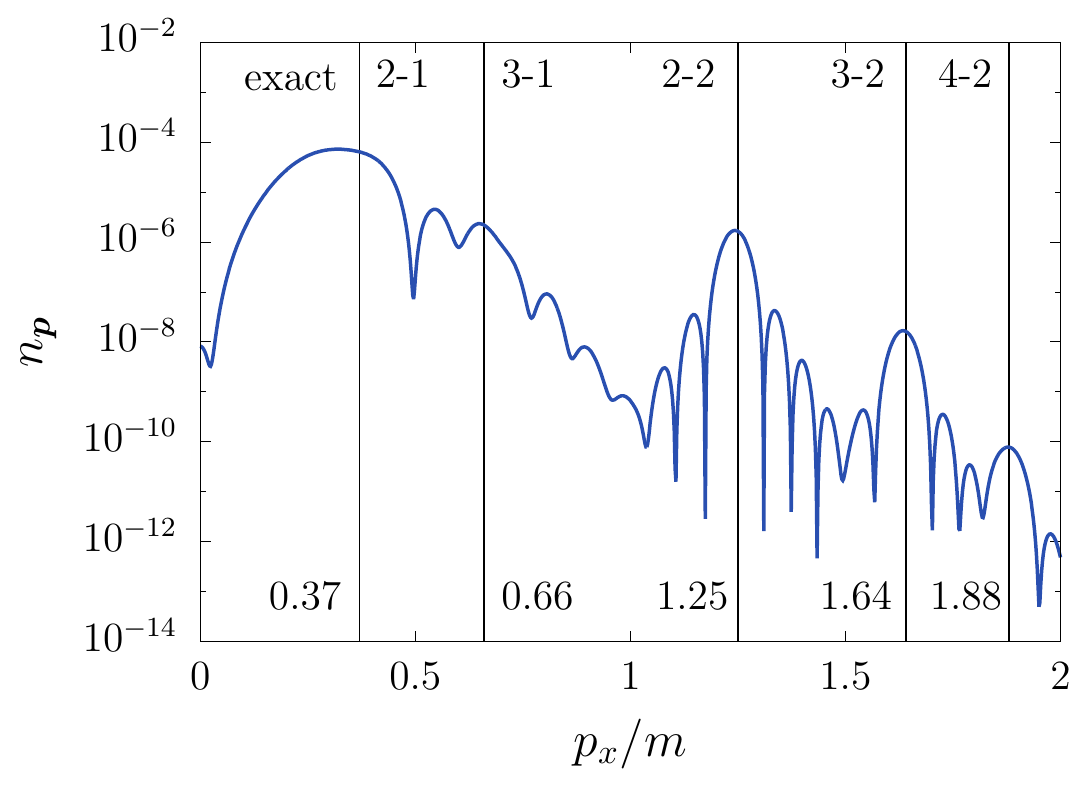}}
\caption{Log-plot of the particle momentum spectrum $n_{\boldsymbol p}$ as a function of $p_x$ for $p_y = p_z = 0$ evaluated within the LDA (left) and calculated exactly for $k_z = \omega$ (right). Vertical lines indicate predictions given by our absorption model. Only for symmetric photon absorption ($2-2$ at $p_{x}=1.25m$) the peak location is unaffected by having a nonzero $k_z$. Every other spike in the spectrum either changes position or splits into multiple smaller peaks. Overall, the particle yield decreases. The field parameters are $\tau = 40 m^{-1}$, $\varepsilon = 0.2$, and $\omega = 0.8m$.}
\label{fig:long_LDA_px}
\end{figure*}

To illustrate our goal, we display in Fig.~\ref{fig:long_LDA_px} an example of the momentum distribution in the direction parallel to the employed electric field (pulse duration $\tau = 40 m^{-1}$, peak field strength $\varepsilon = 0.2$, and temporal frequency $\omega = 0.8m$). We have chosen this direction specifically because it serves as a good starting point for the following discussion on more complex, fully developed 2D spectra. In particular, because although the background fields' spatial frequency still plays a vital role in creating distinctive patterns in the particle distribution, the net momentum transfer in the direction of $p_x$ is always zero leading to an easier to interpret distribution function.

In the left panel of Fig.~\ref{fig:long_LDA_px}, we present the results obtained within the LDA. For comparison, in the right panel we have displayed the momentum spectrum for on-shell photons, where we have $k_z = \omega = 0.8m$. It becomes immediately obvious, that the LDA predictions do not match the exact results when the spatial frequency is properly taken into account. None of the peaks in the spectrum are entirely unaffected by a change in $k_z$. Even for four-photon absorption in the $4$ and $2-2$ processes, the peak height is significantly different.

Concerning the computational time, we first note that our numerical technique based on the Furry-picture formalism allows us to independently calculate any individual points in the spectrum of particles, which makes this method particularly efficient when computing one-dimensional distributions. For instance, the spectrum displayed in Fig.~\ref{fig:long_LDA_px} (right) was obtained with a standard consumer laptop in 4 minutes with the resolution $\Delta p_x = 0.002m$ and at least 5 accurate significant digits (the uncertainty in $n_{\boldsymbol{p}}$ was always less than $10^{-14}$). The LDA performs even faster: it took only one minute to calculate an analogous curve.

The DHW formalism is well suited to perform calculations with respect to the full momentum spectrum. As a matter of fact, in order to obtain the spectrum shown in Fig.~\ref{fig:long_LDA_px} (right) we calculated
the particle momenta in $p_x$--$p_z$ and then cut along $p_z=0$. This, of course, significantly increases the overall CPU time. For a reasonable comparison we therefore divide by the number of grid points in $N_{p_z}=1280$ leading to roughly $10$ minutes of runtime.

Within the LDA, the DHW formalism is converted from a set of partial differential equations to a set of ordinary differential equations. Consequently, the distribution function $n_{\boldsymbol p}$ can be calculated at any point in ${\boldsymbol p}$ individually dramatically improving the overall performance. For example, the curve in Fig.~\ref{fig:long_LDA_px} (left)
has been obtained within $2$ minutes.

\paragraph{Absorption model.---}
The key to understanding this apparent mismatch and multiphoton pair production in general lies in proper application of conservation laws; here for energy and linear momentum (it was already successfully employed, e.g., in Refs.~\cite{aleksandrov_prd_2018, ruf_prl_2009,peng_arxiv_2018}). To derive a conclusive picture, we start by stating the full 4-momentum conservation for any single multiphoton creation process,
\begin{equation}
p_{e^+}^\mu + p_{e^-}^\mu = n_+ k_+^\mu + n_- k_-^\mu, \label{eq:Econs}
\end{equation}      
where $p_{e^-}^\mu$ and $p_{e^+}^\mu$ give the 4-momentum of the electron and positron, respectively. The photons are characterized by the 4-vectors $k_+^\mu$ and $k_-^\mu$, where we used a subscript to distinguish photons with positive and negative momentum with respect to the $z$ axis, cf. the definition of the vector potential~\eqref{eq:field_config}.
In particular, we have
\begin{eqnarray}
 p_{e^+}^\mu = \begin{pmatrix} E_{e^+} \\ p_{x,e^+} \\ p_{y,e^+} \\ p_{z,e^+} \end{pmatrix} ,\quad  p_{e^-}^\mu = \begin{pmatrix} E_{e^-} \\ p_{x,e^-} \\ p_{y,e^-} \\ p_{z,e^-} \end{pmatrix} ,\quad
  k_+ = \begin{pmatrix} \omega \\ 0 \\ 0 \\ k_z \end{pmatrix} ,\quad k_- = \begin{pmatrix} \omega \\ 0 \\ 0 \\ -k_z \end{pmatrix}.
\end{eqnarray}
As one can see, the only relevant parameters in our analysis are the fields' temporal and spatial frequencies, which coincide with the photon energy $\omega$ and photon momentum $k_z$.\footnote{There are of course additional effects to consider if, e.g., the field strength reaches almost critical values or the pulse length is too short. These modifications, however, do not play a significant role in the setups discussed here and, in order to keep the discussion as simple as possible, are thus left out in the main body of the manuscript. We have added some additional remarks on this issue in Appendix~\ref{sec:appendix_effmass}.}

We see immediately, that linear momentum conservation for two components is trivial: $p_{x,e^+}=-p_{x,e^-}$ and $p_{y,e^+}=-p_{y,e^-}$. The momentum $p_z$, however, yields a fascinating connection between electrons, positrons, and photons. Formally, we have 
\begin{equation}
 p_{z,e^+} + p_{z,e^-} = n_+ k_z - n_- k_z.
\end{equation}
In a fully symmetric process, $n_+ = n_-$, the particle-antiparticle pair is created with zero net momentum as the linear momentum of the electron and positron cancel each other out, $p_{z,e^+} = - p_{z,e^-}$. If, however, $n_+ \neq n_-$, then the particle and antiparticle are accelerated in opposite directions minus an offset $q_0$ due to a surplus of momentum stemming from the absorbed photons,
\begin{equation}
q_0 = \frac{k_z}{2} \left(n_+ - n_- \right). \label{eq:q0}       
\end{equation}
The interesting aspect of this relation is that although for the latter 
$p_{z,e^+} + p_{z,e^-} \neq 0$ holds, the whole system's total linear momentum ${\boldsymbol P}_{\rm tot}=\sum_{\rm particles} {\boldsymbol p}$ is still conserved. This is due to the fact that, in order to obtain the full picture we always have to take into account all multiphoton channels, in particular the mirrored case
$n_+ \leftrightarrow n_-$. As the sum over all particle momenta in the combined situation vanishes, the total momentum is preserved (due to vacuum conditions, initially we have $\boldsymbol{P}_{\rm tot}=0$). Nevertheless, to keep the discussion concise, we derive all expressions for a single scattering channel and ignore the mirrored cases.

Particle and antiparticle momenta in the $z$ direction are therefore given by
\begin{eqnarray}
    p_{z,e^+} &=& q_{z} + q_0, \\
    p_{z,e^-} &=& -q_{z} + q_0.
\end{eqnarray}
In the weak-field limit, $\varepsilon \ll 1$, we can assume that the particle energies are given by
\begin{eqnarray}
 E_{e^+} &=& \sqrt{ m^2+ p_x^2 + p_y^2 + (q_z + q_0)^2 }, \label{eq:Ep} \\
 E_{e^-} &=& \sqrt{ m^2+ p_x^2 + p_y^2 + (q_z - q_0)^2 }. \label{eq:Em}
\end{eqnarray}
The final particle momenta for a specific setup can thus be calculated by finding a solution of the equation
\begin{equation}
    \sqrt{ m^2+ p_x^2 + p_y^2 + (q_z + q_0)^2 } + \sqrt{ m^2+ p_x^2 + p_y^2 + (q_z - q_0)^2 } = n_+ \omega + n_- \omega. \label{eq:EnFull}
\end{equation}

\begin{figure}[t]
      \begin{center}
	\includegraphics[height=0.42\linewidth]{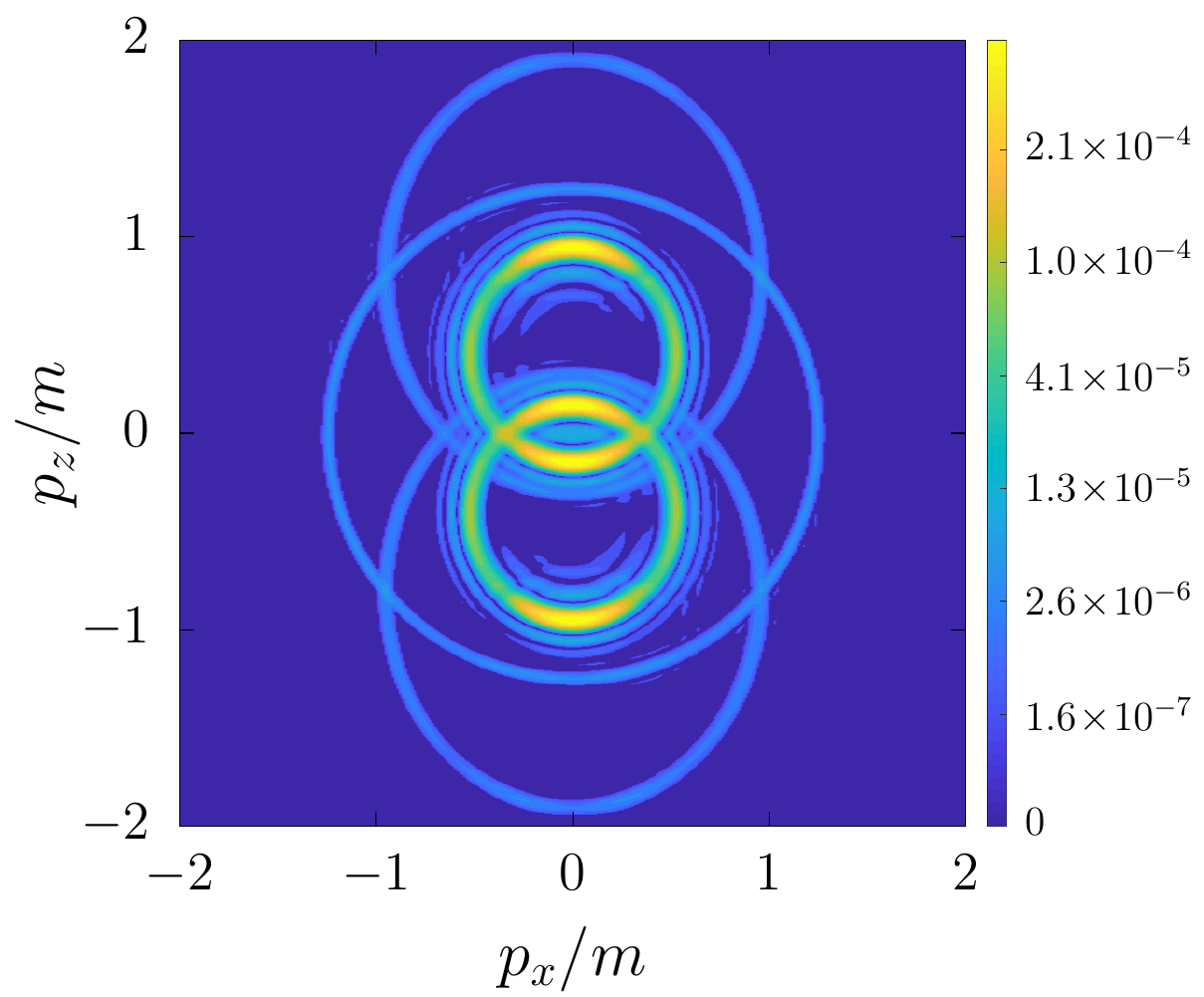}~~~~~\includegraphics[height=0.42\linewidth]{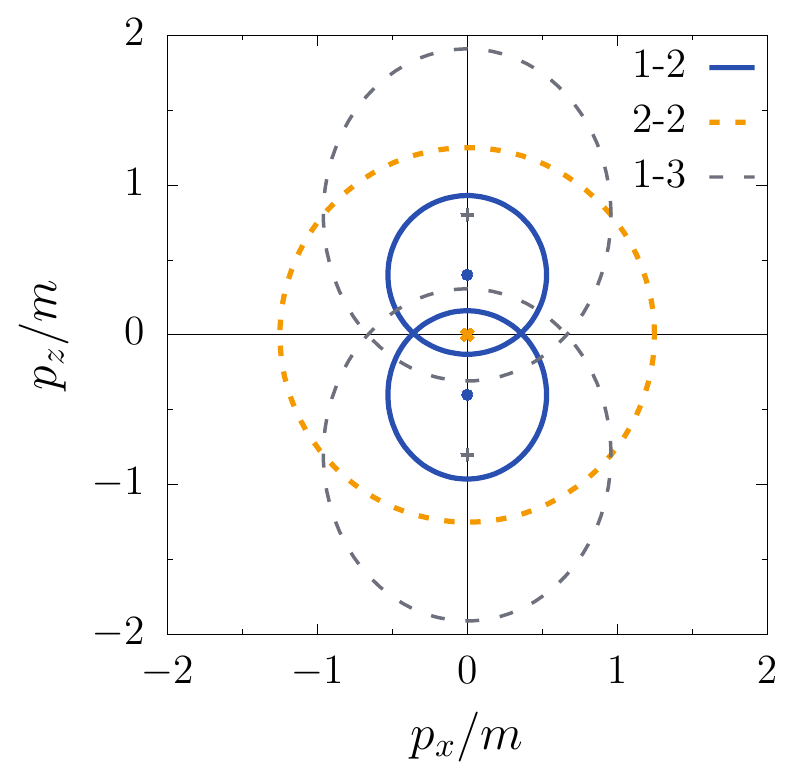} 
      \end{center}
      \caption{Modified density plot of the particle momentum spectrum $n \left( p_x, p_z \right)$ ($p_y = 0$). In order to highlight the weaker production channels, the distribution function is displayed using nonlinear scaling. A sketch of the various open channels calculated through our model is illustrated on the right-hand side.
      The channels $2-1$ and $3-1$ are not labeled explicitly. Field parameters: $\varepsilon =0.2$, $\tau=60m^{-1}$, and $\omega=k_z=0.8m$.} 
      \label{fig:spec_ATI}
      \end{figure}

\paragraph{Scattering channels.---}
Having established such a simple picture of multiphoton pair production, we now turn our attention towards understanding scattering channels; see Fig.~\ref{fig:spec_ATI} for an example. At first we have to establish a general concept of channel openings and the consequences of having a photon energy surplus. Due to the fact that an electron-positron pair has to be
created out of the vacuum by absorbing photons in the first place, the energy of the incident photons combined has to exceed a threshold 
given by the rest mass of the pair, $2m$. If this is not the case, this particular scattering channel is closed.

If, on the other hand, a channel is open, particles are created with a specific kinetic energy. As energy is conserved throughout the creation
process, any excessive photon energy is converted into higher particle momenta by forming above-threshold peaks. In order to analyze these peaks in the particle spectrum, we derive simple expressions within the LDA ($k_z \to 0$) and in the on-shell limit ($k_z = \omega$), respectively. 

\paragraph{Particle spectra.---}
In order to allow for a gentle start, we begin the discussion on the basis of one-dimensional spectra.
For the special case where $p_y=p_z=0$ we obtain in the case of a vanishing spatial frequency $k_z$
\begin{equation}
p_x(k_z=0) = \sqrt{\left[ \frac{ \left(n_+ + n_- \right) \omega}{2} \right]^2 - m^2}
\label{eq:res_px_0}
\end{equation}
recovering the simple energy-momentum relation.
Consequently, within the LDA the positions of the above-threshold peaks are determined solely by the total number of photons absorbed $n = n_+ + n_-$ as there is no distinction between $n_+$ and $n_-$
due to the fact that one cannot distinguish between photons propagating in direction $+z$ and photons propagating in direction $-z$. 

As Fig.~\ref{fig:long_LDA_px} has been the showcase for this section, we apply our analysis to these plots first. From the data we retrieve peak positions at $p_{x}=0.65m$, $1.24m$, and $1.73m$ for the LDA. Employing our absorption model \eqref{eq:res_px_0}, we find that these peaks correspond to our model's predictions regarding three-, four-, and five-photon pair production, respectively. Hence, we label the peak positions $p_{x,3}=0.66m$, $p_{x,4}=1.25m$, and $p_{x,5}=1.73m$.

\begin{figure}[t]
\center{\includegraphics[width=0.65\linewidth]{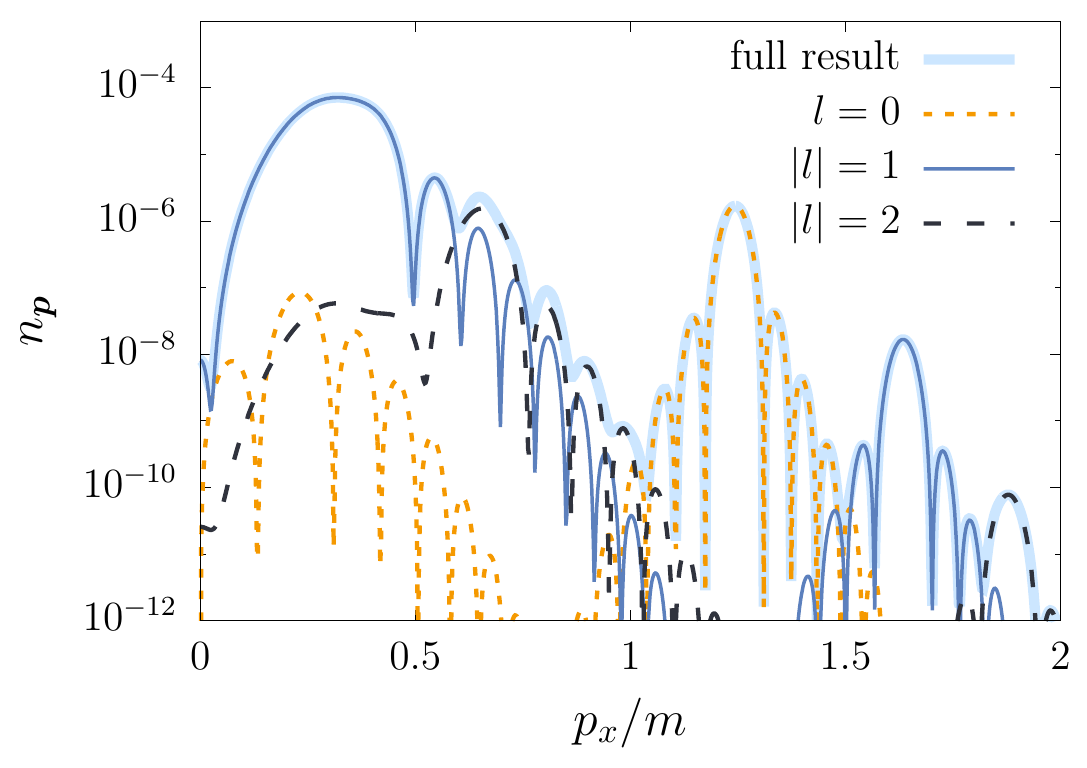}}
\caption{Particle spectrum calculated for a standing wave with temporal length $\tau = 40 m^{-1}$, peak field strength $\varepsilon = 0.2$, and frequency $k_z = \omega = 0.8m$ ($p_y = p_z = 0$). Within the Furry picture, it is possible to discriminate between contributions from different net photon absorption numbers $l=n_+ - n_-$. The thick light-blue line gives the total production rate serving as a reference value.}
\label{fig:long_LDA_px2}
\end{figure}

In the same vein we can perform an analysis for on-shell photons. To be more specific, for $k_z = \omega$ we obtain
\begin{equation}
p_x(k_z=\omega) = \sqrt{\frac{4n_+^2 n_-^2 \omega^2}{(n_+ + n_-)^2} - m^2}
\label{eq:res_px_omega}
\end{equation}
finally allowing us to analyze also the right panel in Fig.~\ref{fig:long_LDA_px}. From Eq. \eqref{eq:res_px_omega} we immediately see that the total number of absorbed photons $n=n_+ + n_-$ is not sufficient anymore to interpret the particle spectrum. In turn, this automatically explains the rather involved spectrum in Fig.~\ref{fig:long_LDA_px} as the various channels become fully distinct. One consequence is that only signatures at high momenta become fully distinguishable. This includes 
the symmetric channel $n_+=n_-=2$, which is still located at the same position $p_{x,2-2}=p_{x,4}=1.25m$ due to a vanishing linear momentum surplus. Additionally, we have the channel $n_+=3,~ n_-=2$ at $p_{x,3-2}=1.64m$ and the channel $4-2$ at $p_{x,4-2}=1.88m$. The broad particle distribution at low momenta $p_x$, however, is the result of summing over various overlapping contributions. In Fig.~\ref{fig:long_LDA_px2} we illustrate the particle spectrum with a focus on the net photon number $l=n_+-n_-$. Apparently, at small $p_x$ the main fraction of particles is created through the channels $2-1$ and $3-1$. Moreover, the distribution function for the channel $4-1$ peaks around $p_x \sim 0.79m$ enhancing the local particle rates even further. Interestingly, there is already evidence for the channel $1-1$ to open at $p_x < 0.5m$ although two-photon pair production is well below the threshold $2 \times 0.8m < 2m$.  


Furthermore, the right-hand side in Eq.~\eqref{eq:res_px_omega} yields only imaginary, and thus unphysical values, for one-sided photon absorption. In other words, multiphoton pair production is generally forbidden if the photon is on shell $\omega=k_z$ and either $n_+$ or $n_-$ vanishes. This is in accordance with the fact that an individual plane-wave pulse cannot produce pairs. In particular, one-photon pair production is therefore only possible if the photon is off shell~\cite{Lv:2018wpn}. 

On a similar note we can derive analytical expressions for the two different limits in the $z$ component of the final particle momentum. For on-shell photons, $k_z=\omega$, our model yields
\begin{equation}
 p_z(k_z=\omega) = \frac{\left(n_+ - n_- \right) \omega}{2} \pm \frac{n_+ + n_-}{2} \sqrt{ \omega^2 - \frac{m^2 + p_x^2 + p_y^2}{n_+ n_-} }, \label{eq:qz2}
\end{equation}
which was also found in Ref.~\cite{aleksandrov_prd_2018}.
Similarly to Eq. \eqref{eq:res_px_omega}, the right-hand side diverges for one-sided photon absorption.


On the contrary, in the homogeneous limit, $k_z=0$, photons carry only energy, which, in turn, enables, e.g., one-photon pair production. Analyzing the expression
\begin{equation}
 p_z(k_z=0) = \pm \sqrt{ \left[ \frac{ \left(n_+ + n_- \right) \omega}{2} \right]^2 - (m^2 + p_x^2 + p_y^2) }, \label{eq:qz3}
\end{equation}
we can further deduce that this includes not only one-photon pair production but also channels where only photons from one beam are absorbed. Furthermore, as $\left(n_+ + n_- \right)$ is equal to the total number of absorbed photons $n$, we have recovered the following simple energy conservation relation given in, e.g., Ref.~\cite{kohlfuerst_prl_2014}:
\begin{equation}
 \left( \frac{ n \omega}{2} \right)^2 = m^2 + \boldsymbol p^2.
\end{equation}

The ring pattern revealed in Fig.~\ref{fig:spec_ATI} can be described if we keep both $p_x$ and $p_z$ in Eq.~(\ref{eq:EnFull}). In particular, we obtain the following relation:
\begin{equation}
\frac{4n_+ n_-}{(n_+ + n_-)^2}\, (p_z-q_0) ^2 + p_x^2= n_+ n_- \omega^2 - m^2.
\label{eq:ellipses}
\end{equation}
This equation indicates that each of the ``resonance rings'' is, in fact, an ellipse with semiminor axis $b = \sqrt{n_+ n_- \omega^2 - m^2}$ corresponding to the $p_x$ direction and semimajor axis $a = b (n_+ + n_-)/(2\sqrt{n_+ n_-}) \geq b$ regarding the $p_z$ direction. The center of the ellipse is located at $p_x = 0$, $p_z = q_0$. These predictions of our model are in perfect agreement with the full-simulation results.

\paragraph{Resonances.---}
With the aid of the model introduced above, we can also efficiently search for resonance effects in the spectrum. A similar study has already been performed on the basis of two-level systems exploiting Rabi oscillations in order to determine resonance frequencies with maximal transition probability between the two states~\cite{mocken_pra_2010, akal_prd_2014, aleksandrov_prd_2017_1} (in the case of more complex space-time-dependent setups, it was done in Refs.~\cite{ruf_prl_2009,aleksandrov_prd_2018,woellert_2015}). 
In terms of our model, such a study can be equivalently formulated as a search for optima in the particle distribution for particles at rest $p_x=p_y=p_z=0$. The corresponding optimal frequencies are found to be
\begin{equation}
 \omega_{\boldsymbol {p = 0}}(k_z) = \omega_0(k_z) = \frac{m + \sqrt{m^2 + k_z^2 \left( n_+ - n_- \right)^2 }}{n_+ + n_-}. \label{eq:w0}
\end{equation}
We point out that such resonances precisely at $\boldsymbol{p} = 0$ appear only for odd values of $n = n_+ + n_-$ due to parity and charge-conjugation symmetry~\cite{mocken_pra_2010, akal_prd_2014, aleksandrov_prd_2017_1,aleksandrov_prd_2018}.

\begin{figure}[t]
\center{\includegraphics[width=0.65\linewidth]{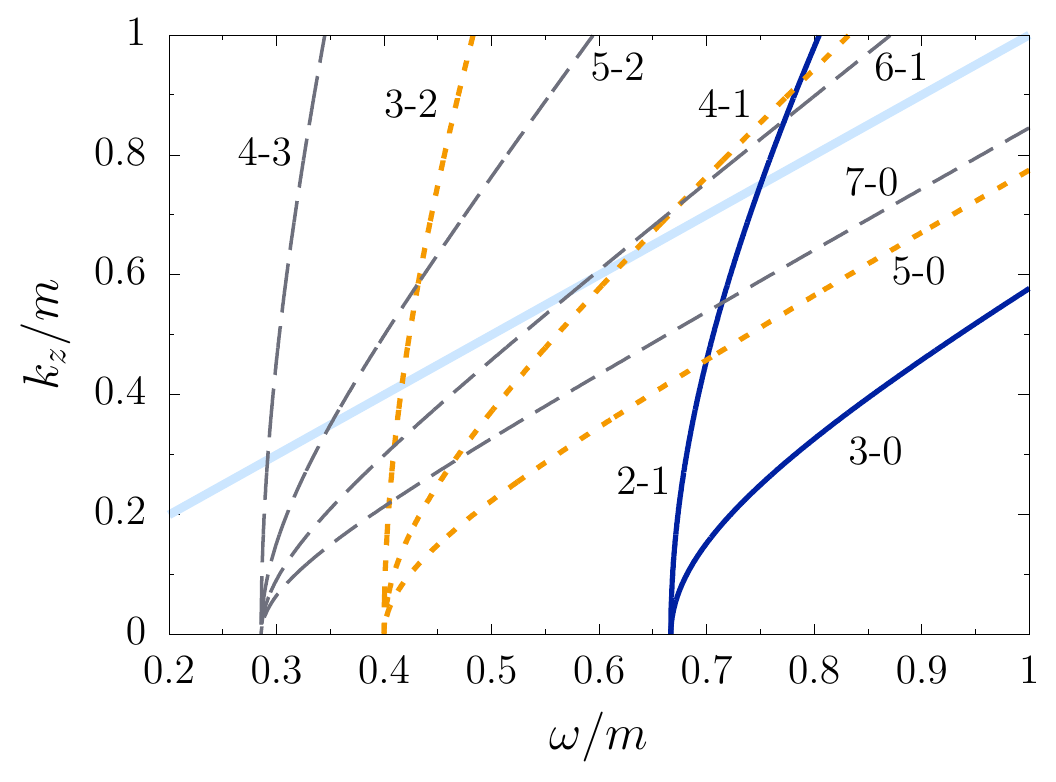}}
\caption{Map of the resonance peaks as a function of temporal $\omega$ and spatial frequency $k_z$ for odd values of the total number $n = n_+ + n_-$ of photons absorbed. The peaks are color coded according to $n$ ($n = 3$, $5$, and $7$).
The thick, solid light-blue line is an indicator for $k_z=\omega$.}
\label{fig:res_omega_peaks}
\end{figure}

To demonstrate conclusively how big the impact of a nonzero spatial frequency is, we illustrate Eq.~\eqref{eq:w0} in Fig.~\ref{fig:res_omega_peaks}. By varying the spatial frequency $k_z$, we can easily calculate the peak positions for any given absorption channel. The resulting 2D map eventually reveals a highly complex structure of peak locations in the multiphoton regime. Furthermore, it makes it very clear that taking into account the spatial dependence of the background field completely changes the landscape of few-photon pair production. Most important, however, it shows spectacularly that within the LDA we obtain nonphysical contributions. Simply because none of the $n_+ - 0$ lines in Fig.~\ref{fig:res_omega_peaks} intersect with the on-shell limiter (light-blue thick line), pair creation, for which either $n_+$ or $n_-$ is zero, is impossible for on-shell photons. On a similar note, for $k_z=0$ multiple above-threshold peaks overlap at the same position in the spectrum leading to nonphysical enhancements. Such an effect can already be observed in Fig.~\ref{fig:long_LDA_px}, where the four-photon channel is misleadingly large in the LDA.

For the sake of completeness, the on-shell limit for this type of resonance is given by
\begin{equation}
 \omega_0(k_z=\omega_0) = \frac{m}{2} \frac{n_+ + n_-}{n_+ n_-}.
\end{equation}
The LDA yields
\begin{equation}
 \omega_0(k_z=0) = \frac{2m}{n_+ + n_-}.
 \label{eq:w0_res0}
\end{equation}
%
%
Keep in mind, though, that these zero-momentum frequencies $\omega_0$ only help to determine spikes in the local pair production rate. They do not yield any information regarding the total particle number $N$. More specifically, only a tiny fraction of electrons/positrons are created at rest at these frequencies as most of the particles are going to be produced with nonvanishing momentum.
Moreover, the frequencies $\omega_0$ are not directly linked to the resonance frequencies $\omega_{\ast}$
and thus to the sudden increases in the production rate as the latter are expected to happen when a new channel opens for $p_x=p_y=q_z=0$. 

In order to find these resonance frequencies $\omega_{\ast}$, we again rely on our model. In this case, energy conservation gives
\begin{equation}
 2 \sqrt{m^2 + q_0^2} = (n_+ + n_-) \omega.
\end{equation}
A resonance in the particle yield is then obtained for frequencies
\begin{equation}
 \omega_{\ast} (k_z) = \sqrt{ \frac{4m^2 + k_z^2 \left(n_+ - n_- \right)^2 }{\left( n_+ + n_- \right)^2} }, \label{eq:om}
\end{equation}
and, for the sake of completeness, the resonance condition for on-shell photons is given by
\begin{equation}
 \omega_{\ast}(k_z=\omega_{\ast}) = \frac{m}{\sqrt{n_+ n_-}}.
\end{equation}
Additionally, for light that does not carry momentum, Eq.~\eqref{eq:om} takes on the more familiar form
\begin{equation}
 \omega_{\ast}(k_z=0) = \frac{2 m}{n_+ + n_-}.
 \label{eq:w_res0}
\end{equation}
Hence, only within the LDA does the appearance of a resonance in the particle spectrum for particles at rest coincide with the resonance condition for the particle yield, cf. Eq.~\eqref{eq:w0_res0} and Eq.~\eqref{eq:w_res0}.

\begin{figure}[t]
      \begin{center}
      \includegraphics[height=0.38\linewidth]{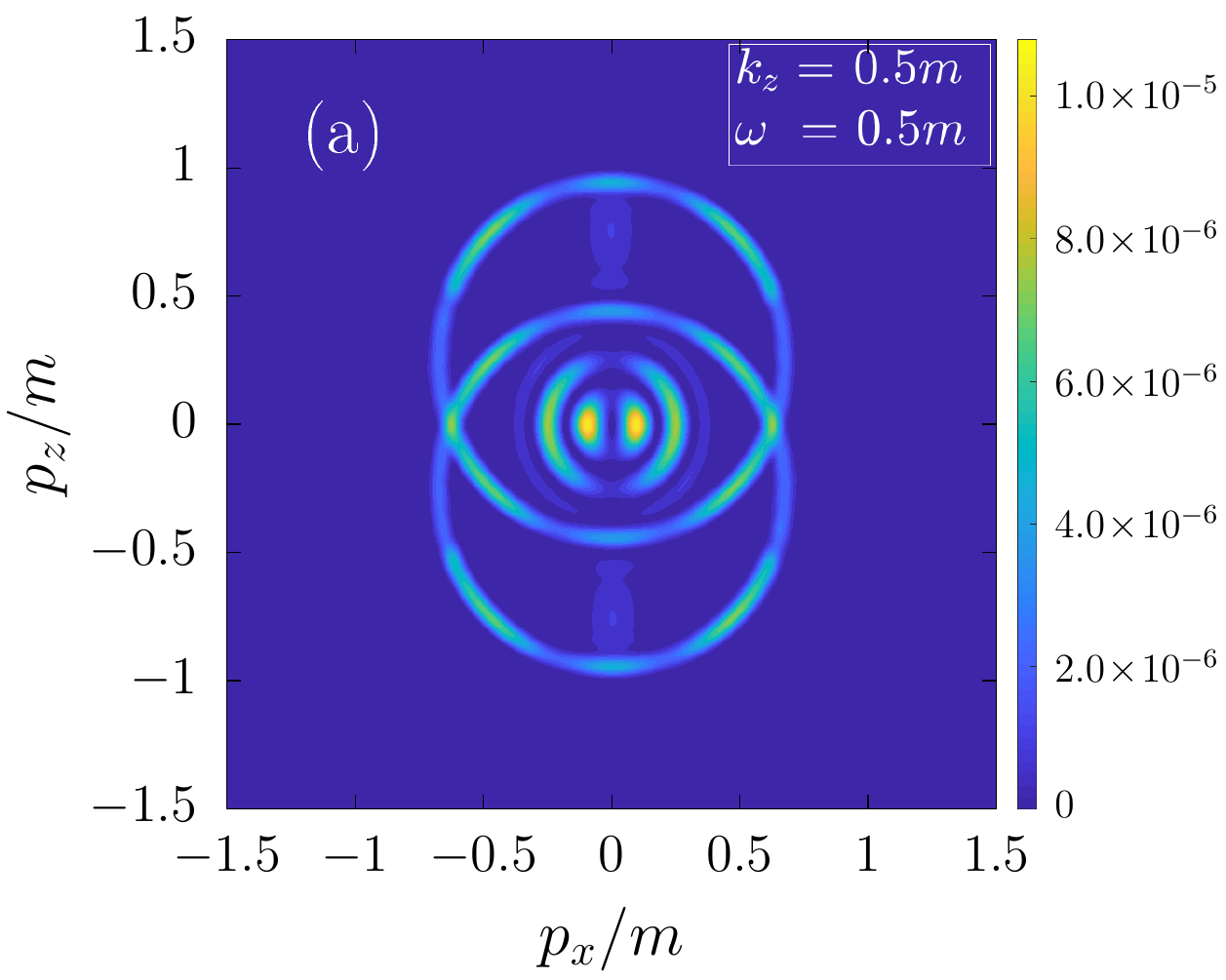}~~~~~\includegraphics[height=0.38\linewidth]{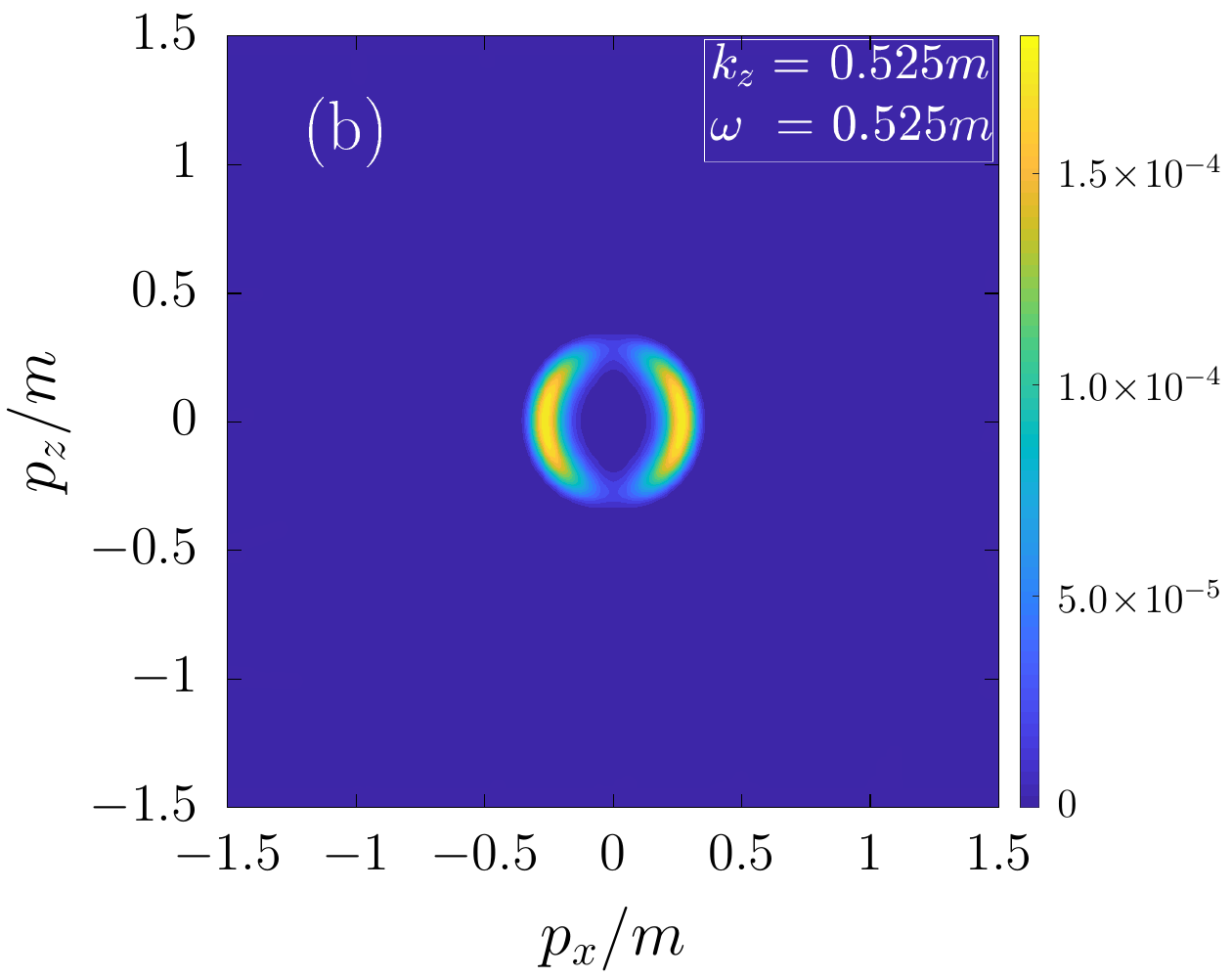} \\ \vspace{0.015\linewidth}
      \includegraphics[height=0.38\linewidth]{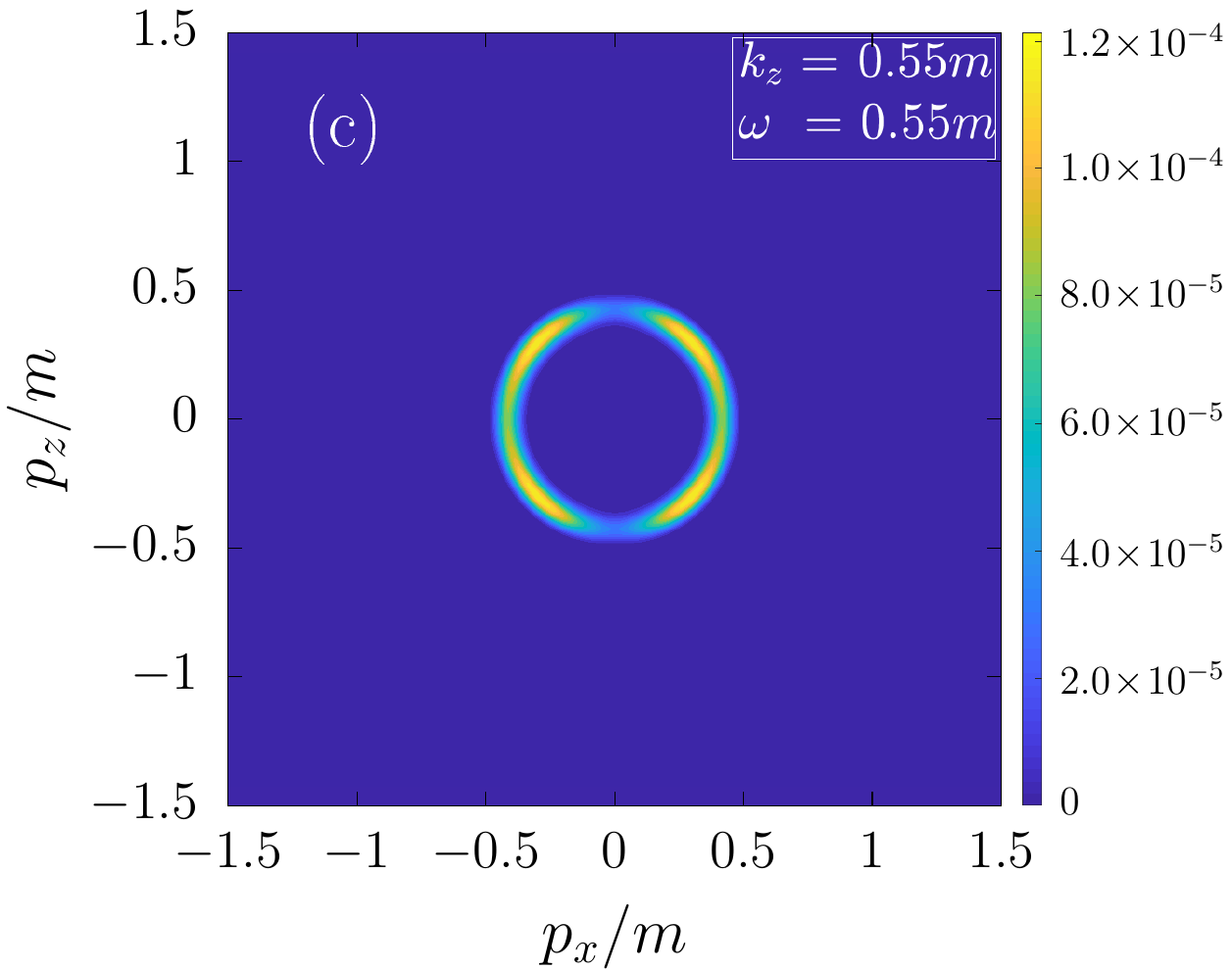}~~~~~\includegraphics[height=0.38\linewidth]{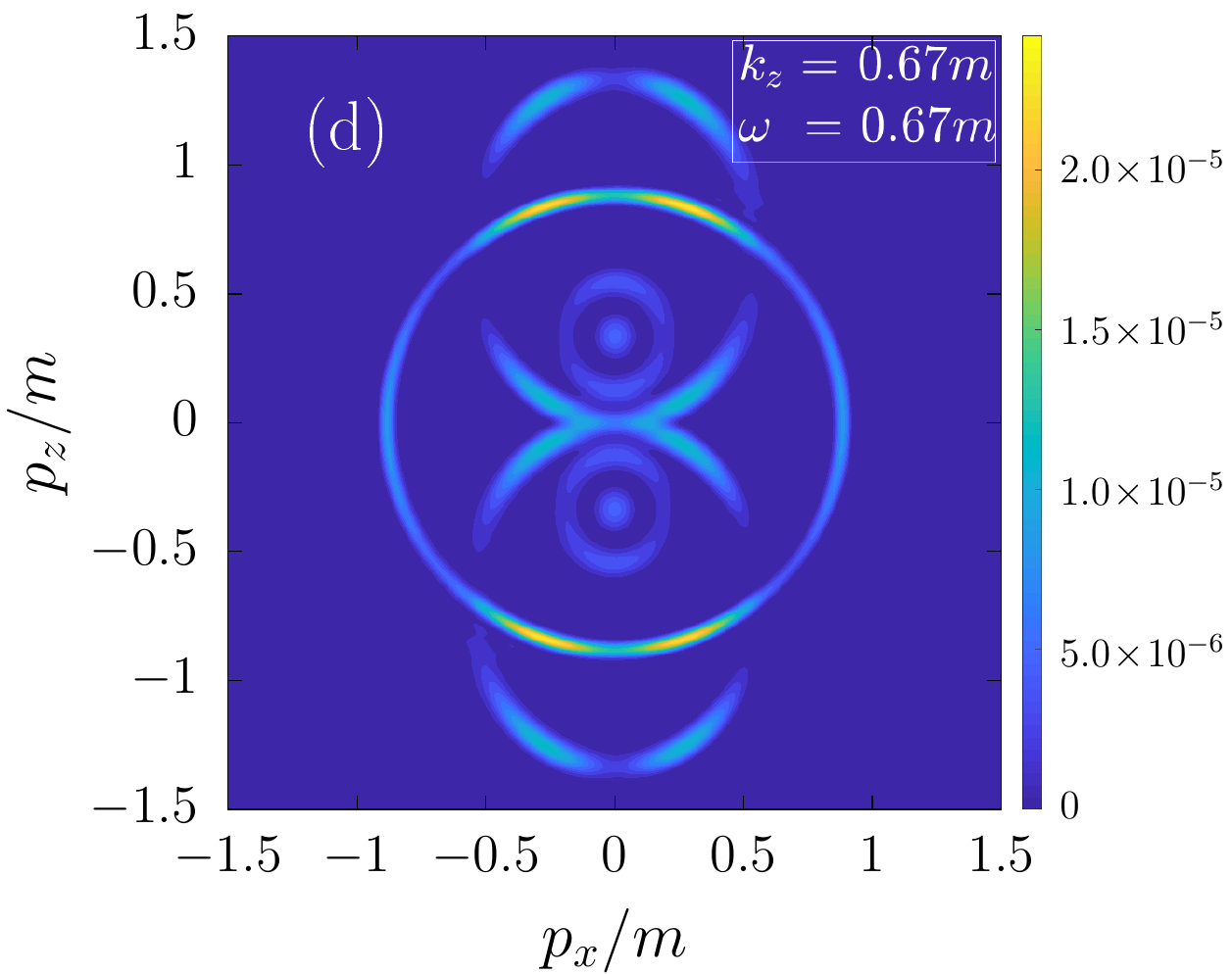}
      \end{center}
      \caption{Density plots of the particle momentum spectrum $n \left( p_x, p_z \right)$ ($p_y = 0$) for fields of different temporal $\omega$ and spatial $k_z$ frequency 
      with a peak field strength of $\varepsilon =0.2$ and a pulse duration of $\tau=60m^{-1}$. The symmetric four-photon channel ($n_+=n_-=2$) is about to open at $\omega=k_z=0.5m$ 
      [double peak at $p_z=0$ in panel (a)]. The channels $3-2$ and $2-3$ contribute heavily to
      the particle yield at $\omega=k_z=0.5m$ \big[ellipses with center at $\boldsymbol{p} = \left( 0, \, 0.25m \right)$\big]. Their impact is overshadowed by contributions from four-photon channels once the threshold is reached. At $\omega=k_z=0.67m$, the channels $2-2$ 
      (ring with center at the origin), $1-3$, and $3-1$ are clearly visible. Additionally, around $p_z = \pm 0.335m$, signatures of three-photon pair production are already observable.} 
      \label{fig:spec_omega}
      \end{figure}

In Fig.~\ref{fig:spec_omega} we display this connection for a series of setups. Most notably, in Fig. \ref{fig:spec_omega}(d) three different channels are observable (not counting the two mirrored cases). The channels $2-3$ and $3-2$ qualify for resonance in the zero-momentum frequency $\omega_0$. However, the majority of particles created within this channel are ejected at a $\sim 30^\circ$ angle from the center. Additionally, although $\omega_0$ exhibits a local optimum due to the $3-2$ channel, the actual resonance in the particle yield $N \left( \omega_{\ast} \right)$ is expected to come from the $2-1$ channel. In Fig.~\ref{fig:spec_omega}(d) at $p_z \approx \pm 0.335m$ one can already see that a new peak forms.

\begin{figure}[t]
      \begin{center}
    \includegraphics[height=0.33\linewidth]{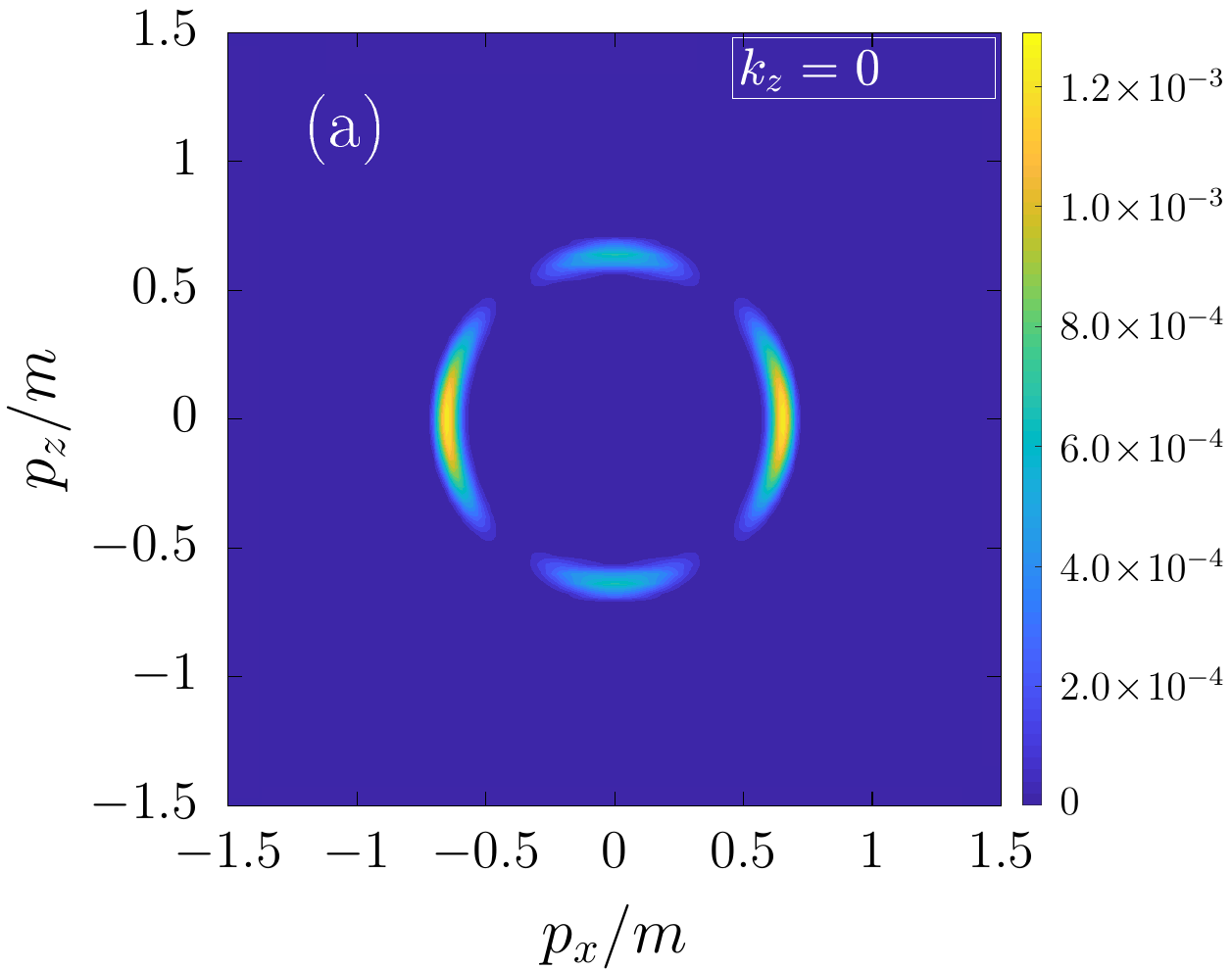}~~~~~\includegraphics[height=0.33\linewidth]{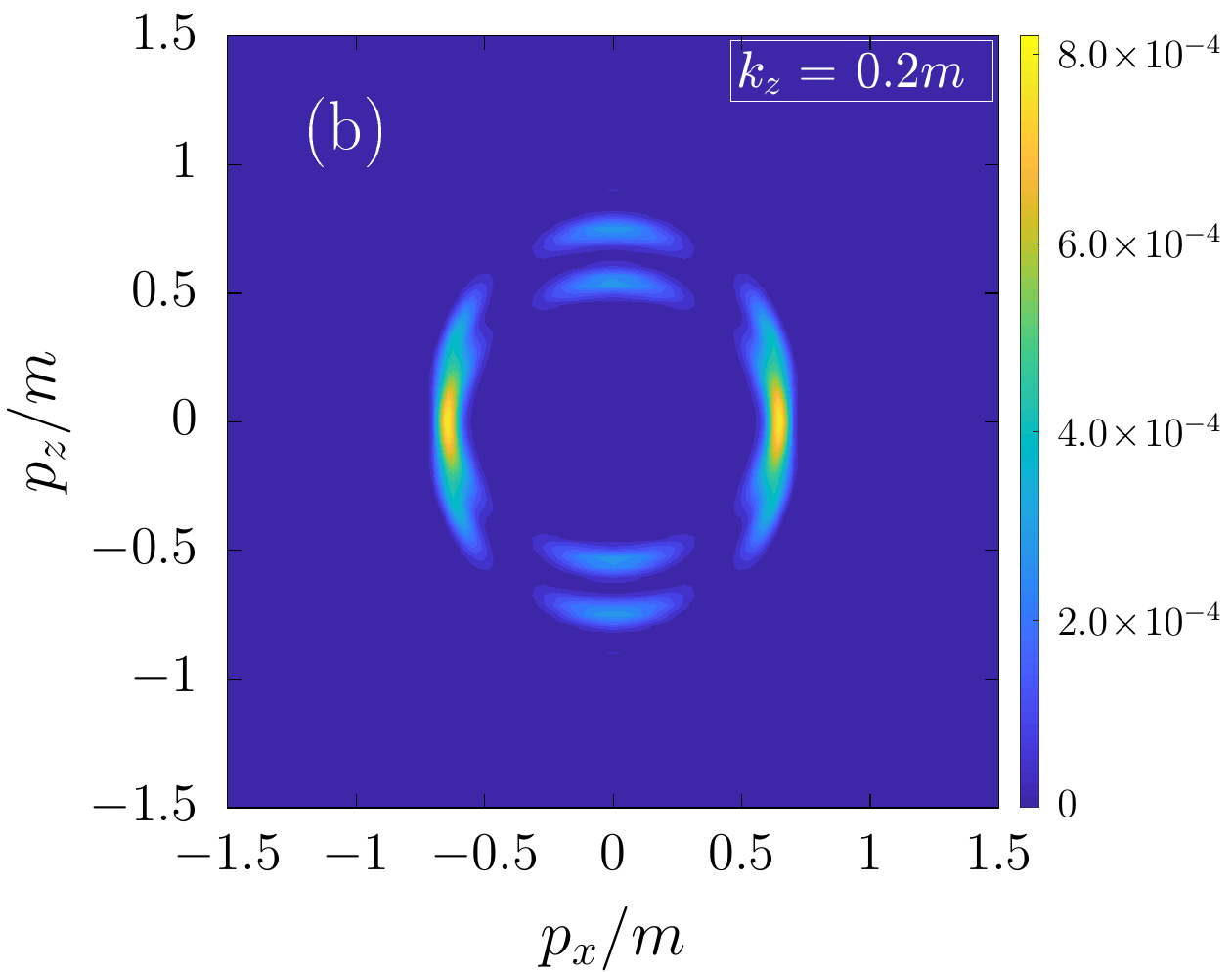} \\ \vspace{0.015\linewidth}
    \includegraphics[height=0.33\linewidth]{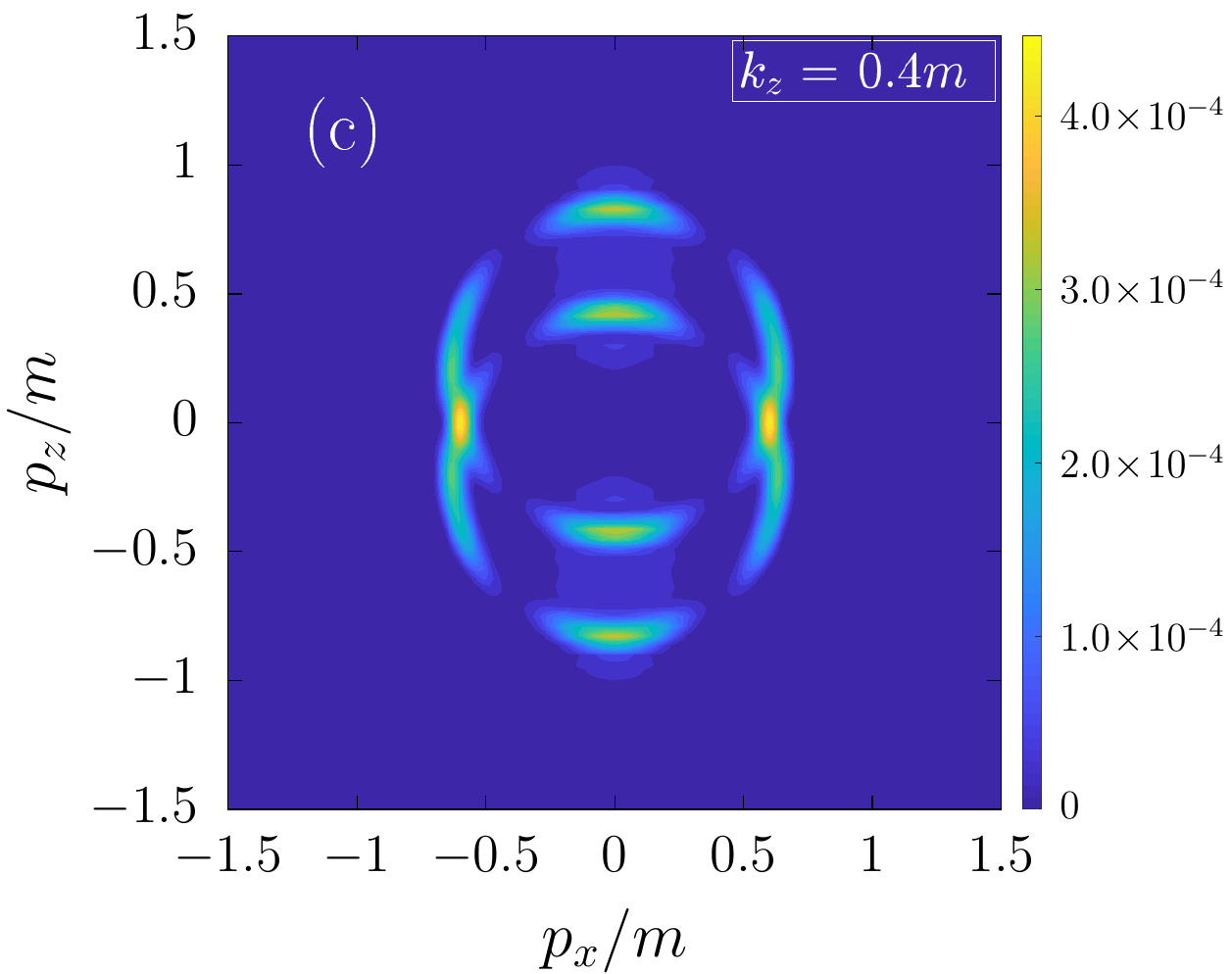}~~~~~\includegraphics[height=0.33\linewidth]{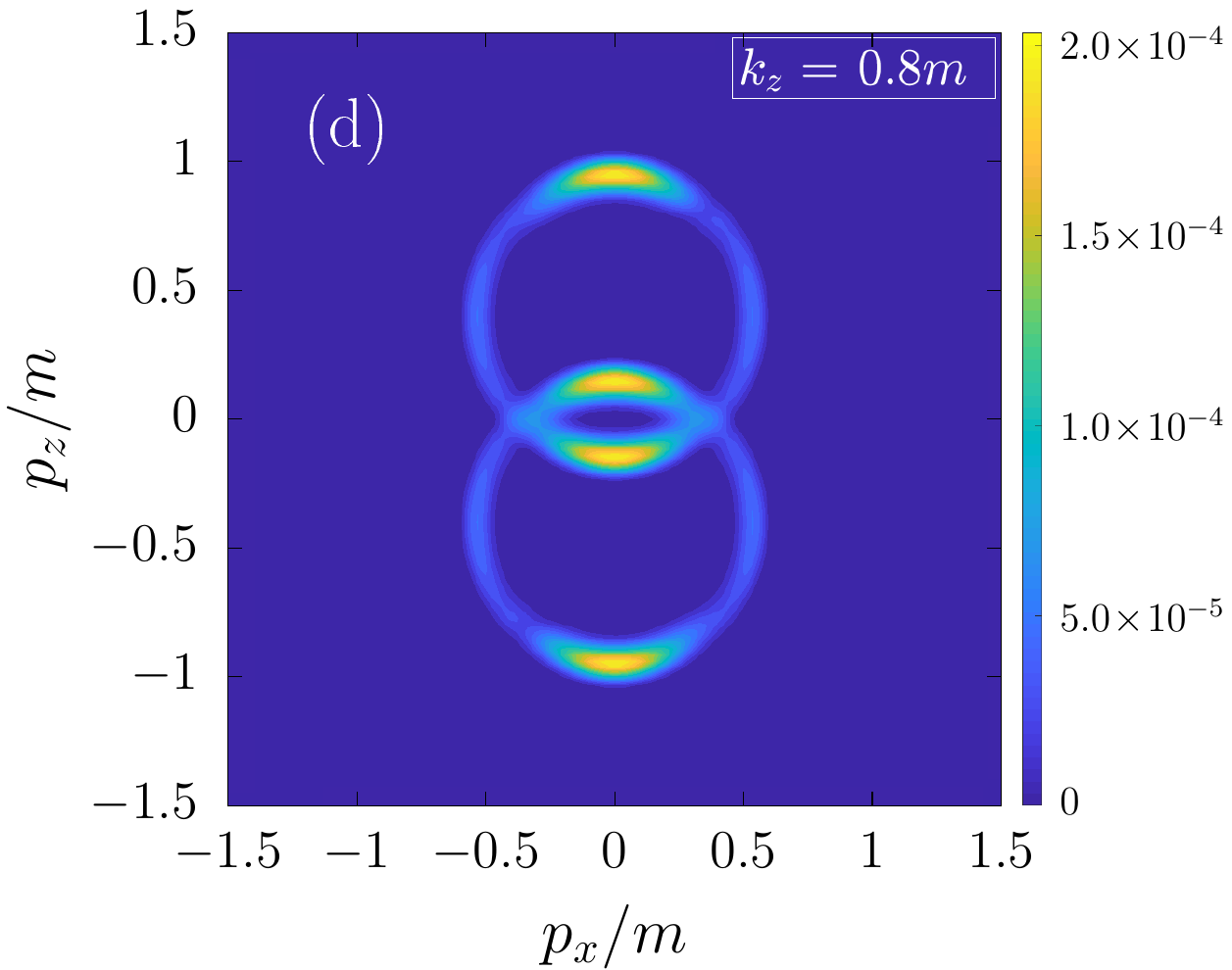} \\ \vspace{0.015\linewidth}
    \includegraphics[height=0.33\linewidth]{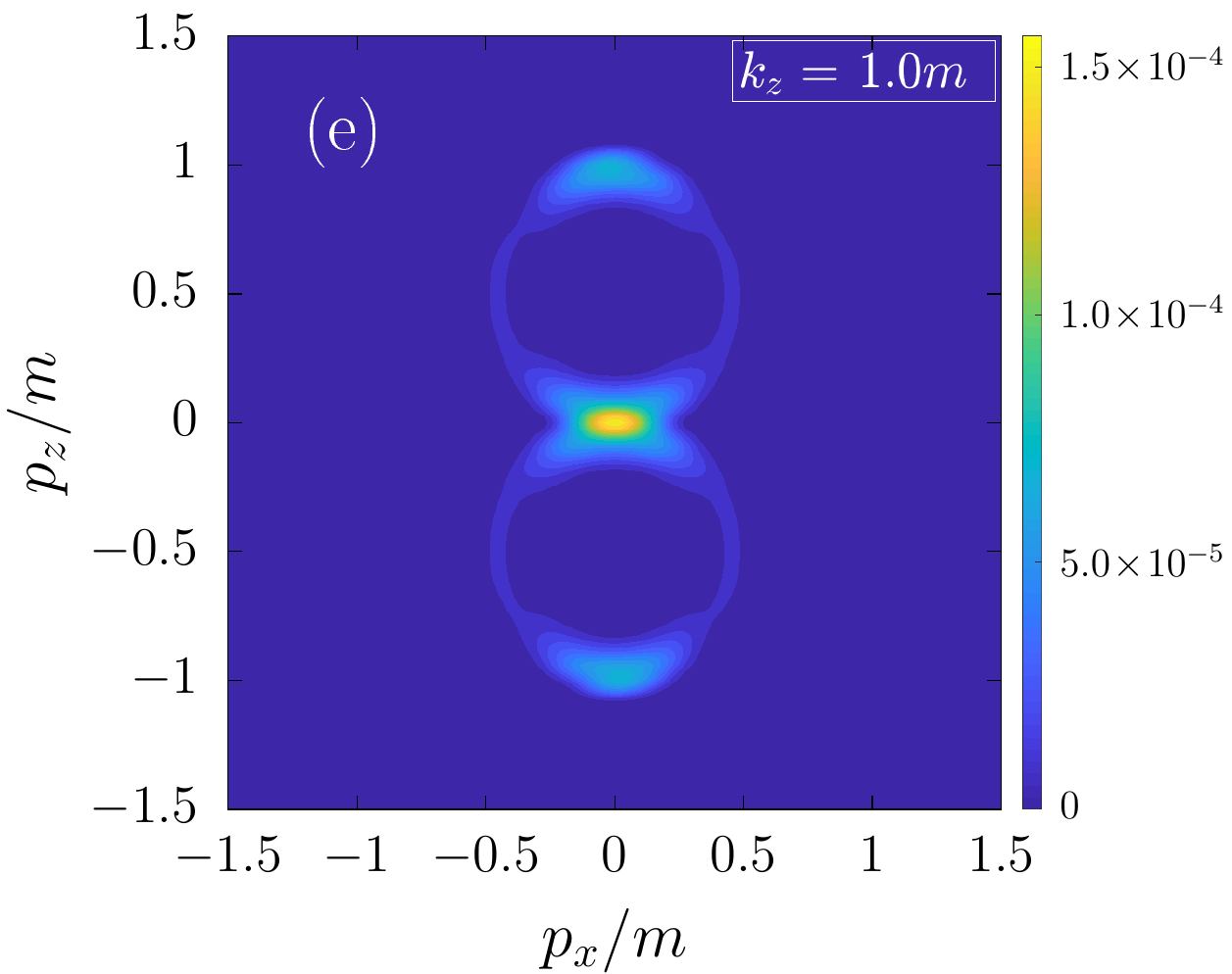}~~~~~\includegraphics[height=0.33\linewidth]{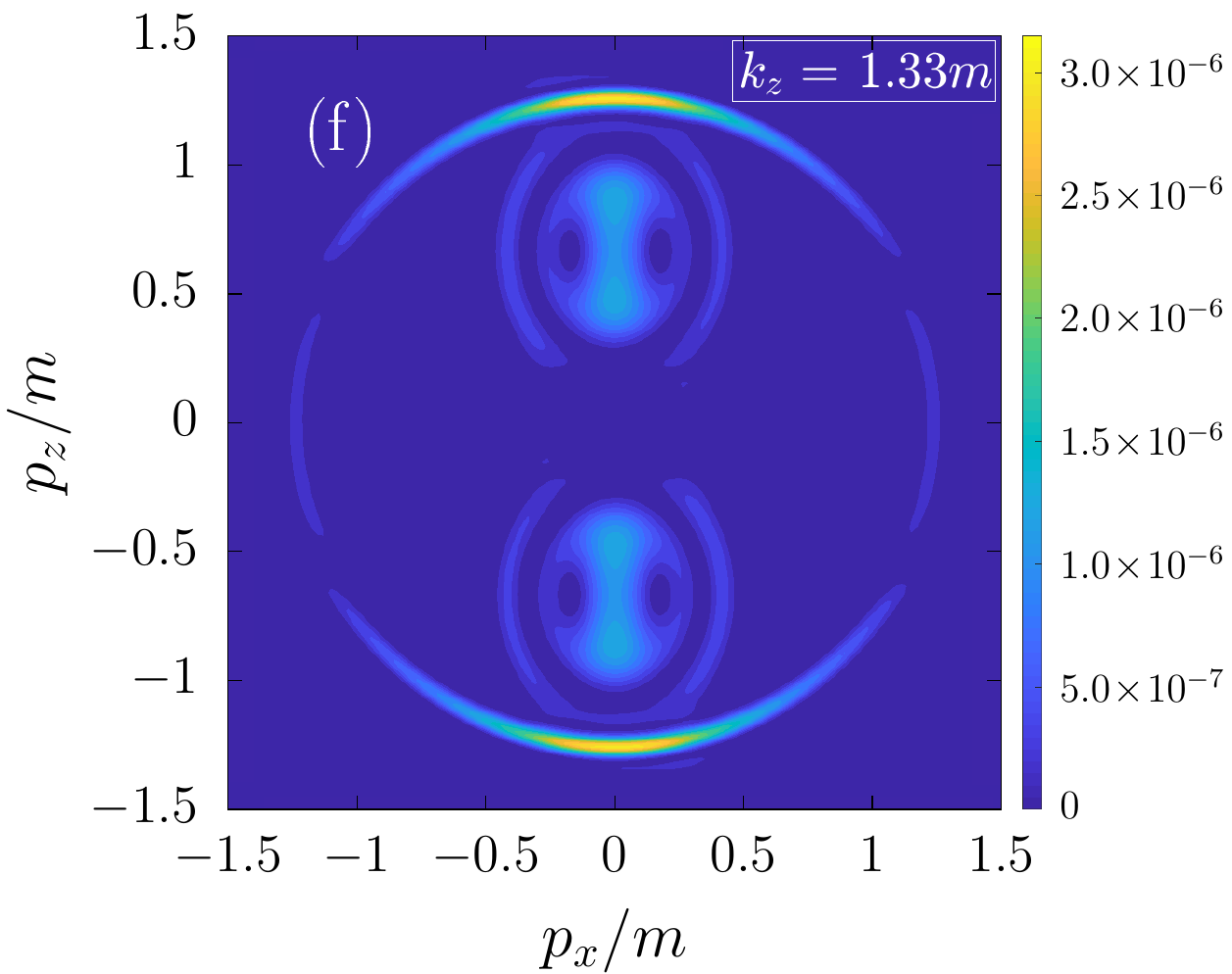}
      \end{center}
      \caption{Density plots of the particle momentum spectrum $n \left( p_x, p_z \right)$ ($p_y = 0$) for various spatial frequencies $k_z$. For otherwise identical field
      configurations ($\varepsilon =0.2$, $\tau=40m^{-1}$, $\omega=0.8m$), a higher frequency $k_z$ results in a shift towards higher particle momenta. Going from
      low to high frequencies, the signatures also become smaller in diameter. At a photon momentum of $k_z = 1.0m$, a great amount of particles is created at rest, $\boldsymbol{p}=0$. In panel (f) ($k_z=1.33m$), the channels $2-1$ and $1-2$ are close to the resonance condition. Additionally, the interference patterns are very sensitive to changes in $k_z$; see the shift in the particle distribution's maxima.} 
      \label{fig:spec_k}
      \end{figure}

Last, one fascinating feature of our setup is that it allows one to study channel closing directly. Solving Eq. \eqref{eq:om} for $k_z$ we obtain the resonance condition for the spatial frequency 
\begin{equation}
 k_{z,\ast} = \sqrt{ \frac{\left( n_+ + n_- \right)^2  \omega^2 - 4m^2} {\left( n_+ - n_- \right)^2} }.
\end{equation}
In Fig. \ref{fig:spec_k} we have displayed the full particle momentum spectrum for a configuration with $\varepsilon=0.2$, $\tau=40m^{-1}$, $\omega=0.8m$, and various $k_z$. This series of plots strikingly demonstrates the transition from the LDA results to the on-shell limit up to the value of $k_z$ where the resonance condition is met. 

Technically, for vanishing $k_z$ we cannot distinguish between the $2-1$ and $1-2$ channels. Consequently, both of them form the same structure in momentum space centered at $\boldsymbol {p=0}$ even constructively interfering with each other. We also see, that an increase in the spatial frequency automatically leads to an increase in the net momentum transfer as $n_+ \neq n_-$. Closing of the channels $2-1$ and $1-2$ is predicted at $k_{\ast} \approx 1.33m$ explaining the tremendous decrease in the distribution function in Fig.~\ref{fig:spec_k}(f). The drop-off is apparently so significant that it leads to the $2-2$ channel being dominant. 

\paragraph{Beyond the weak-field approximation.---}
In this whole section, we have never mentioned shifts in the final particle spectrum due to the oscillatory energy of the created particles. The reason is that for the exemplary setups presented here such a field-dependent effect amounts to a $\sim 1 \%$ correction only. Nevertheless, in order to build a bridge to the established results given in the literature we have included a discussion on the effective mass effects in Appendix~\ref{sec:appendix_effmass}.

\clearpage

\section{Discussion and Summary}\label{sec:discussion}

Electron-positron pair production within high-intensity electromagnetic background fields is a highly complex process, and thus adequately precise theoretical predictions for realistic scenarios are difficult to obtain. Specifically, the fact that phenomena concerning the particle distributions occur within laser fields requires resolving effects on two vastly different scales ($1/m \sim 10^{-21}~\text{s}$ and $1~\text{fs} \sim 10^{-15}~\text{s}$), so a complete simulation is out of reach for the foreseeable future.  In this regard and especially with a view to upcoming strong-field laser facilities, it is thus of utter importance to develop a formalism that can accurately describe such strong-field experiments. 

In this work we have therefore inspected the applicability of the LDA in the context of short-pulsed fields and in the regime of few-photon pair production. Within our study, we have found a second criterion for the applicability of the LDA based on the expected particle movement within the background field, the first involving the formation length of the particle pair. To be more specific, in order for the LDA to hold, the particle trajectories after creation must be confined to an area smaller than the field's wavelength. Note, however, that neither the formation length nor the particle movement make  hard, exclusive statements as we have still found setups where only one of these two criteria was fulfilled.

In the second part of the manuscript, we made a comparison between the LDA and full simulation in the regime of multiphoton pair production on the basis of the particle momentum spectrum. We reproduced the spectra given in the literature for the LDA confirming that these momentum distribution functions indeed carry signatures of particle absorption. However, upon further inspection, these distributions fail to give the right above-threshold peak positions (except for the symmetric case, where an equal amount of photons was absorbed from each direction) and even show nonphysical contributions. Moreover, quantum resonances show up at wrong frequencies and momenta and, worse, the amplitudes in the spectrum are wrong by at least 1 order of magnitude. The latter is true even for symmetric, few-photon signals, where enhancements stemming from quantum interferences of different scattering channels are rare.

All limitations of the LDA in this regime can be traced back to its inability to take into account the photon's linear momentum. As a matter of fact we have developed an effective model based upon the conservation of energy and linear momentum in order to predict peak locations and to obtain the right resonance conditions. While this model does not yield any information on the signal strength, it can give an idea about the landscape in the multiphoton regime. This includes estimates for quantum resonances, channel-resolved peak locations as well as zero-momentum resonance frequencies. Although the model presented in this manuscript is based on monochromatic, linearly polarized fields, the basic concept is universal and can be readily applied to more involved field configurations.

\section*{Acknowledgments}

We want to thank Ralf Sch\"utzhold for his support. I.~A.~A. acknowledges the support from the Russian Foundation for Basic Research (RFBR) (Grant No. 20-52-12017), Saint Petersburg State University (SPbSU) (Grant No. 11.65.41.2017), and the Foundation for the advancement of theoretical physics and mathematics ``BASIS.'' The work of C.~K. was initially funded by the Bundesministerium f\"ur Bildung und Forschung (BMBF) under grant No.~05P15SJFAA (FAIR-APPA-SPARC) and by the
Helmholtz Association through the Helmholtz Postdoc Programme (PD-316). This research was supported in part through computational resources provided
by the Helmholtz Institute Jena and the University Jena through the Helmholtz Postdoc Programme (PD-316).

\appendix
\section{Computational methods}\label{sec:appendix}

Within the Furry-picture formalism, the ODE system~(\ref{eq:w_eq_gen}) is solved by means of various Runge-Kutta methods including, for instance, the implicit Gauss-Legendre method of order six. In our computations the explicit schemes turn out to be always stable, so the usage of implicit ones is not necessary. At each time step, we make sure that the components of $w^j_{\boldsymbol{p},s} (t)$ with large values of $|j|$ do not contribute, i.e., the ``momentum box'' is sufficiently wide.

Solution techniques as well as the computational methods used in order to solve the DHW equations \eqref{eq_3_1}--\eqref{eqn1_4} have been already presented in great detail
in Refs.~\cite{kohlfuerst_epjp_2018} and \cite{Kohlfurst2019}. Regarding the computer libraries in use, in order to perform the time integration, we rely on a Dormand-Prince Runge-Kutta integrator
of order 8(5,3)~\cite{NR}. Spatial and momentum derivatives have been carried out with FFTW3~\cite{Frigo05thedesign} using Ref.~\cite{Boyd} as a manual.

A main concern when discussing results on the basis of two quite different numerical approaches is given by the uncertainty as well as the stability of the results. As no proof demonstrating the equality of both techniques nor an analytical solution in the regimes of interest to gauge the methods is available, we can only provide evidence that both formalisms contain the same physics and are indeed equal in terms of producing the correct outcome.

Optimally, different formalisms contain the same information; only the way to retrieve it might be different. Within this manuscript, we employ one solution technique based on Furry-picture quantization uniquely suited to calculate particle distributions to an unprecedented level of accuracy. The alternative way is given by the DHW formalism, a relativistic quantum kinetic approach, that automatically yields, by its design, a complete overview of the particle phase space and therefore the full momentum spectrum.

As demonstrated in the manuscript multiple times, the strengths of these two significantly different formalisms lie in totally different areas: one is about high precision while the other aims at completeness. It is therefore challenging to show that both approaches indeed produce the same result. In this appendix, we want to provide some insight into the technical details of our computations. To do so, we employ a test scenario that allows us to examine in particular the accuracy of the results displayed in the main text.

\begin{figure}[t]
      \begin{center}
      \includegraphics[width=0.65\linewidth]{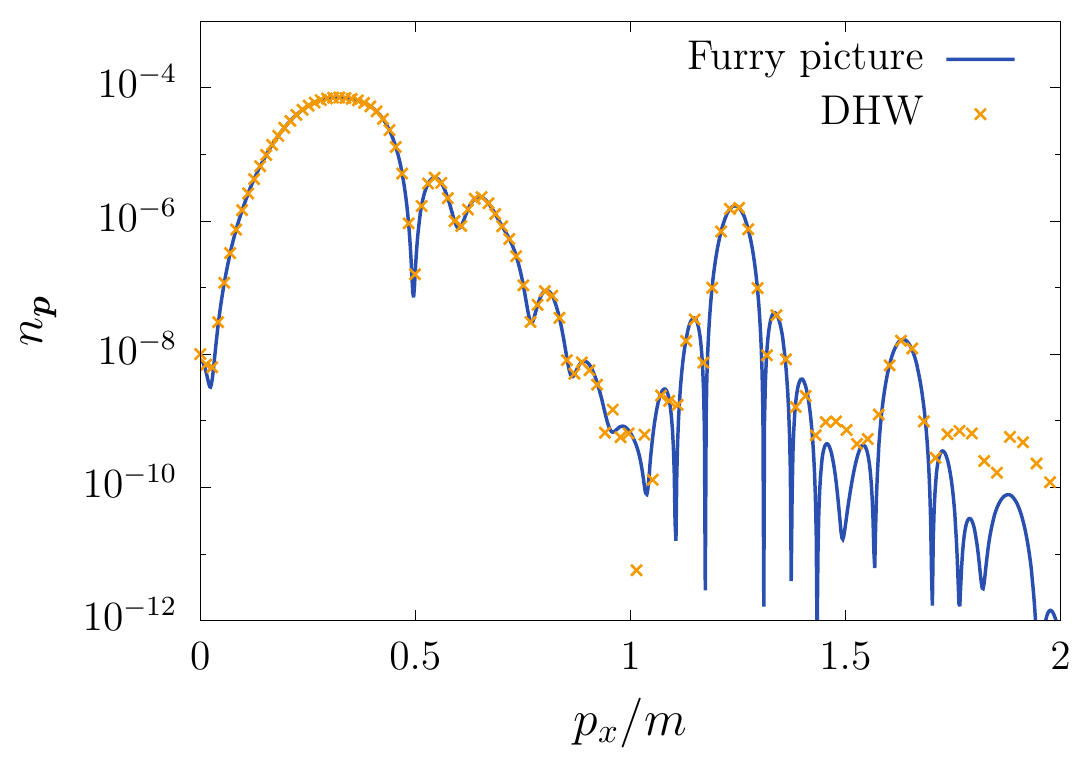} 
      \end{center}
      \caption{Comparison of the particle distribution $n_{\boldsymbol p}$ as a function of the momentum $p_x$ ($p_y = p_z = 0$) obtained through the Furry-picture quantization approach (blue line) and DHW formalism (orange markers). 
      At a peak value around $p_x = 0.32m$, both approaches deviate by roughly $3$\%. The configuration tested here is $\tau=40^{-1}m$, $\varepsilon=0.2$, $\omega=k_z=0.8m$. The uncertainty in $n_{\boldsymbol{p}}$ in the Furry-picture formalism is less than $\sim 10^{-14}$. The DHW formalism, on the other hand, is only accurate to the percentage level plus having a general threshold in uncertainty of $\sim 10^{-9}$ due to the necessity of performing a vast number of complex calculations.}
      \label{fig:app_comp}
      \end{figure}

\begin{figure}[t]
      \begin{center}
      \includegraphics[width=0.65\linewidth]{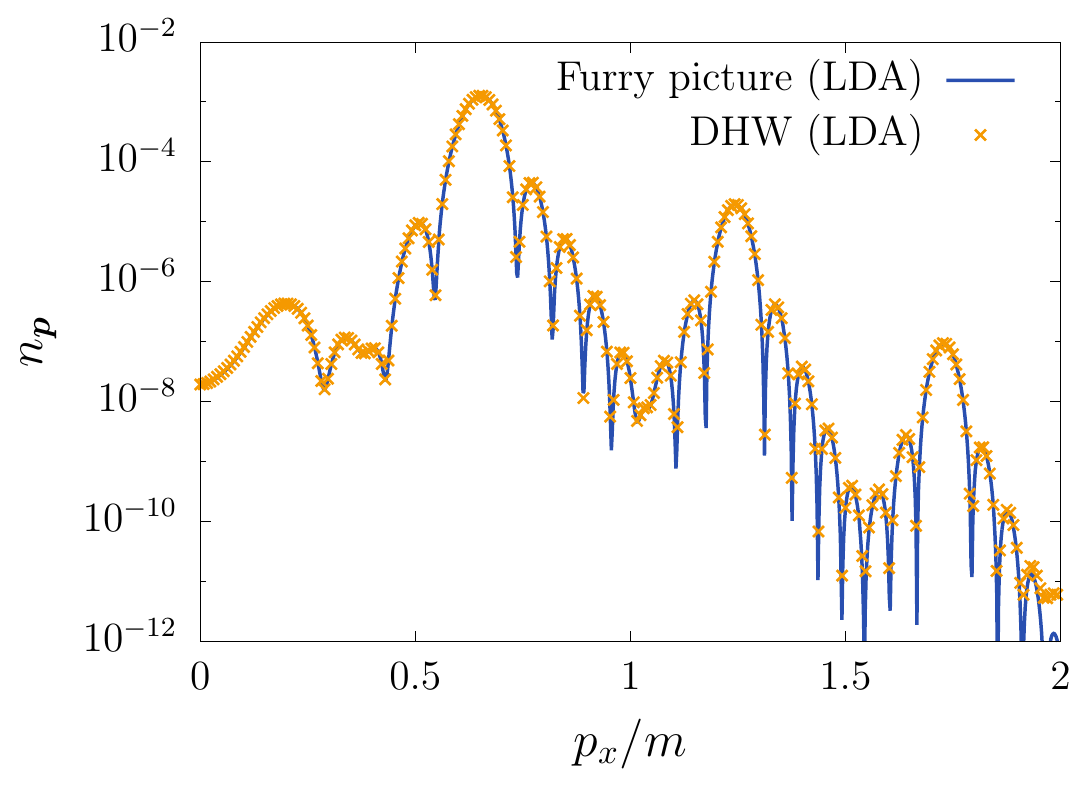} 
      \end{center}
      \caption{Comparison between Furry-picture quantization and DHW formalism in the LDA on the basis of the particle distribution function $n_{p_x,p_y=0,p_z=0}$. Both approaches produce results that are accurate to at least 8 orders. The difference at peak values in $n_{\boldsymbol p}$ is less than $3.5$\%. Field parameters: $\tau=40^{-1}m$, $\varepsilon=0.2$, $\omega=k_z=0.8m$.} 
      \label{fig:app_compLDA}
      \end{figure}

The test configuration is given by a long-pulsed field with the pulse length $\tau=40^{-1}m$, peak field strength $\varepsilon=0.2$, and temporal frequency $\omega=0.8m$. For the spatial frequency we opted for two different scenarios. In the first setup, we studied on-shell photons, $k_z=\omega=0.8m$, ultimately proving the trustworthiness of both approaches; see Fig. \ref{fig:app_comp}. In the second scenario, cf. Fig. \ref{fig:app_compLDA}, we have averaged over $z$ allowing for inspection within the LDA. 

The Furry-picture formalism is formulated in terms of solving many first-order, ordinary differential equations in time. As memory is not an issue, the only concern is about resolving any given oscillation in the mode functions.  As the approach is used sparsely in the manuscript mainly in order to determine distribution functions locally, there is no need to use coarse step functions in order to save computation time. Consequently, the usage of a small step size in time $t$ leads to tremendously accurate solutions in the particle momentum spectra. In Figs.~\ref{fig:app_comp} and \ref{fig:app_compLDA} the actual limiter is given by the precision of long double variables $\sim \mathcal O \left(10^{-15} \right)$. 
The DHW approach, on the other hand, operates intrinsically with partial differential equations. Hence, the accuracy and precision of such an approach cannot compete with more streamlined formalisms using, e.g., mode decomposition. However, given sufficiently advanced computational techniques and sophisticated parallelization schemes, phase-space approaches can make up for their slowness in calculating distributions at a specific point in the domain by the huge amount of output they can produce. On top of that, although kinetic approaches are not on the same level as other formalisms in terms of accuracy, they still provide reasonably precise results, in Fig. \ref{fig:app_comp} the orange curve evens out at $\sim \mathcal O \left(10^{-9} \right)$.

Within the LDA the situation changes significantly. Due to the fact that within this approximation, spatial inhomogeneities are singled out altogether, the coupled system of partial differential equations governing the particle dynamics in the DHW formalism is converted into a set of ordinary differential equations. Consequently, the time integration of this noncoupled system behaves similarly to the mode equations within the Furry-picture formalism leading to highly accurate, easily calculable distribution functions. In Fig. \ref{fig:app_compLDA} a comparison between the Furry-picture formalism and DHW approach in the LDA is displayed.

\section{Effective mass effects}\label{sec:appendix_effmass}

To put it simply, the sole idea of having an effective mass is based on the concept of summarizing the electrons' and positrons' collective interactions with the environment and absorbing it in a single variable. If successful, this approach leads to a massive simplification of the process one wants to describe. It is therefore not surprising that the concept of an effective mass is very often used in the literature when discussing physics in the multiphoton regime. In its most common form, it appears as a field-dependent modification $m \to m_{\ast} \left( \boldsymbol E \right)$ with primary focus on the particles' oscillatory energy within a high-frequency field. 

The huge advantage of this build is its simplicity. In its most uncomplicated form, the effective mass is easily derived through a single averaging over the temporal shape of the vector potential. Reference \cite{kohlfuerst_prl_2014} is an excellent example of such an approach, where the field-dependent mass term was found to be
\begin{equation}
    m_{\ast} = m \sqrt{1+ \xi^2},~~{\rm where}~~\xi = \frac{e}{m} \sqrt{- \langle A^\mu A_\mu \rangle }.
\end{equation}
For the vector potential in this manuscript~\eqref{eq:field_config}, this would amount to
\begin{equation}
    m_{\ast} \approx m \sqrt{1+ \frac{\varepsilon^2 m^2}{2 \omega^2} } 
\end{equation}
completely neglecting any spatial dependency. 

A slightly different approach was given in Refs.~\cite{ruf_prl_2009,aleksandrov_prd_2018,Otto:2014ssa}, where the effective mass was not calculated directly but through an effective particle energy defined via
\begin{equation}
    {\cal E} = \frac{1}{T} \int_0^T {\rm d}t' \sqrt{m^2 + \left({\boldsymbol p} - e {\boldsymbol A} \right)^2}.
\end{equation}
This technique is more accurate than the previous one and applicable even for very strong fields $\varepsilon \sim {\cal O} \left( 1 \right)$. The downside, however, is given by the fact that in order to calculate peak locations \eqref{eq:res_px_0}--\eqref{eq:qz3} or resonance frequencies, Eqs.~\eqref{eq:w0} and \eqref{eq:om}, one has to solve an implicit equation. 

\section{Supplements}\label{sec:appendix_supp}

We could not put all data we have generated in the main body of the manuscript without seeming to be repetitive. The utility of the following attachments is too narrow to warrant putting it into the main body of the manuscript. Nevertheless, they might still turn out to be of good use.

\begin{table}[b]
\caption{Supplemental material for the figures displayed in Sec.~\ref{sec:results_long}: list of peak locations in the particle spectrum according to the exact numerical simulation $p_x$ and our model $p_{x,n_+-n_-}$, where $n_+$ ($n_-$) gives the number of photons absorbed with positive (negative) momentum and $l=n_+ - n_-$. Effective mass effects are not taken into account in Sec. \ref{sec:results_long}, and thus the peak location for the channel $2-1$ is slightly off. ${}^a$ The background field configuration is determined by a temporal length of $\tau=40 m^{-1}$, a peak field strength of $\varepsilon=0.2$, and a frequency of $k_z=\omega=0.8m$.}
\label{tab:peaks}
\begin{center}
\begin{tabular}{lccrr}
\hline
\hline
 $l$ \hspace{0.5cm} & $n_+$ & $n_-$ & \hspace{0.25cm} $p_{x,n_+-n_-}[m]$ & \hspace{0.25cm} $p_x[m]$  \\
 \hline
 1& 2 & 1 & 0.37 \footnotemark & 0.32 \\  
 2& 3 & 1 & 0.66 & 0.656 \\
 3& 4 & 1 & 0.8 &  0.662-0.794 \footnotemark \\
 0& 2 & 2 & 1.25 & 1.244 \\
 1& 3 & 2 & 1.64 & 1.634 \\
 2& 4 & 2 & 1.88 & 1.878 \\
 3& 5 & 2 & 2.06 & 2.048 \\
 0& 3 & 3 & 2.18 & 2.178 \\
 1& 4 & 3 & 2.55 & 2.55 \\
 2& 5 & 3 & 2.83 & 2.824 \\
 \hline
 \hline
\end{tabular}
\end{center}
${}^a$  {\footnotesize If effective mass effects were taken into account, we would get $p_x=0.325m$.} \\
${}^b$  {\footnotesize This configuration exhibits a plateau rather than a peak.}
\end{table}

      \begin{figure}[t]
      \begin{center}
	\includegraphics[width=0.65\linewidth]{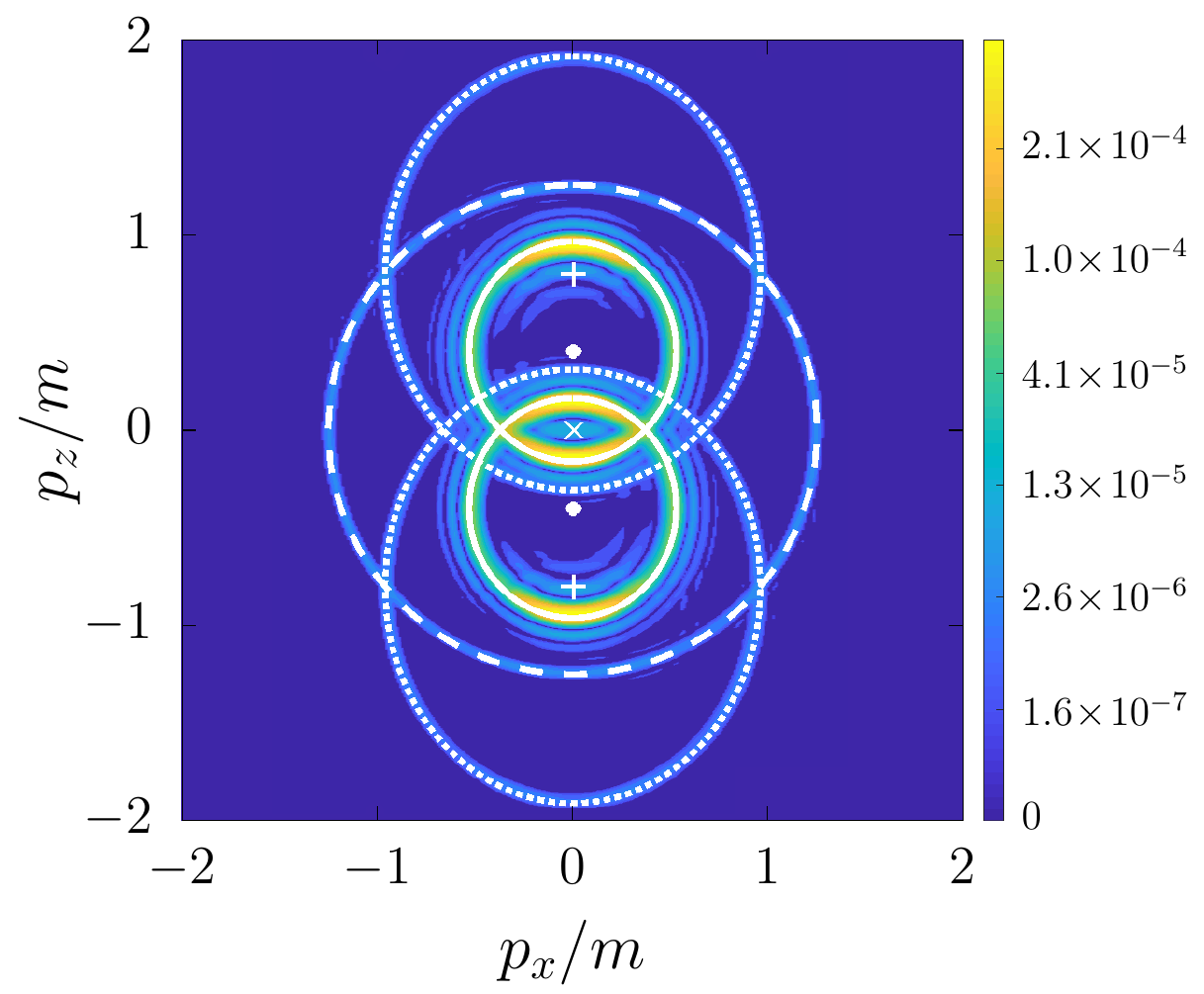} 
      \end{center}
      \caption{Supplemental material for Fig. \ref{fig:spec_ATI}. Overlay showing the particle spectrum obtained from solving the DHW equations in a combination with the solution within our absorption model. The channels $2-2$ (ring with center ``x'' at origin), $1-3$ and $3-1$ (ellipses with center ``$+$'' at $p_z=\pm 0.67m$) are visible as well as $1-2$ and $2-1$ 
      (ellipses with center ``o'' at $p_z=\pm 0.335m$). Field parameters: $\varepsilon =0.2$, $\tau=60m^{-1}$, and $\omega=k_z=0.8m$.} 
      \label{fig:app_spec_ATI}
      \end{figure}     

\clearpage



\begin{thebibliography}{99}
\bibitem{comment_exact} In what follows, by ``exact computations'' we mean calculations performed to zeroth order in the fine-structure constant without any additional approximations regarding the spatiotemporal dependence of the classical external field.
\end{thebibliography}
\end{document}